\documentclass{aa}                    
\usepackage{graphicx}                
\usepackage{txfonts}                 

\newcommand{\Halpha}{H$\alpha$}

\newcommand{\kms}{km\,s$^{-1}$}
\newcommand{\ms}{m\,s$^{-1}$}

\begin{document}

\title{PEPSI deep spectra\thanks{Based on data acquired with PEPSI using the Large Binocular Telescope (LBT) and the Vatican Advanced Technology Telescope (VATT).
The LBT is an international collaboration among institutions in the
United States, Italy, and Germany. LBT Corporation partners are the
University of Arizona on behalf of the Arizona university system; Istituto
Nazionale di Astrofisica, Italy; LBT Beteiligungsgesellschaft, Germany,
representing the Max-Planck Society, the Leibniz-Institute for Astrophysics
Potsdam (AIP), and Heidelberg University; the Ohio State University; and
the Research Corporation, on behalf of the University of Notre Dame,
University of Minnesota and University of Virginia.}}

\subtitle{II. Gaia benchmark stars and other M-K standards}

\author{K. G. Strassmeier, I. Ilyin, \and M. Weber}

\institute{Leibniz-Institute for Astrophysics Potsdam (AIP), An der Sternwarte
    16, D-14482 Potsdam, Germany; \\ \email{kstrassmeier@aip.de},
    \email{ilyin@aip.de}, \email{mweber@aip.de}}

\date{Received ... ; accepted ...}

\abstract{High-resolution \'echelle spectra confine many essential stellar parameters once the data  reach a quality appropriate to constrain the various physical processes that form these spectra.}{We provide a homogeneous library of high-resolution, high-S/N spectra for 48 bright AFGKM stars, some of them approaching the quality of solar-flux spectra. Our sample includes the northern \emph{Gaia} benchmark stars, some solar analogs, and some other bright Morgan-Keenan (M-K) spectral standards.}{Well-exposed deep spectra were created by average-combining individual exposures. The  data-reduction process relies on adaptive selection of parameters by using statistical inference and robust estimators. We employed spectrum synthesis techniques and statistics tools in order to characterize the spectra and give a first quick look at some of the science cases possible.}{With an average spectral resolution of $R\approx 220,000$ (1.36\,\kms ), a continuous wavelength coverage from 383\,nm to 912\,nm, and S/N of between 70:1 for the faintest star in the extreme blue and 6,000:1 for the brightest star in the red, these spectra are now made public for further data mining and analysis. Preliminary results include new stellar parameters for 70~Vir and $\alpha$~Tau, the detection of the rare-earth element dysprosium and the heavy elements uranium, thorium and neodymium in several RGB stars, and the use of the $^{12}$C to $^{13}$C isotope ratio for age-related determinations. We also found Arcturus to exhibit few-percent \ion{Ca}{ii} H\&K and \Halpha\ residual profile changes with respect to the KPNO atlas taken in 1999.}{}


\authorrunning{K. G. Strassmeier et al.}

\titlerunning{PEPSI deep spectra. II. Gaia benchmark stars and other M-K standards}

\maketitle

\section{Introduction}

High-resolution spectroscopy is key for progress in many fields of astrophysics, most obviously for fundamental stellar physics. The advent of ESA's {\sl Gaia} mission spurred a large high-spectral-resolution survey of FGK-type stars at ESO with significant results now coming online (e.g., Smiljanic et al. \cite{smil14}). This survey, dubbed the {\sl Gaia}-ESO survey (Gilmore et al. \cite{gil}), also defined a list of 34 benchmark FGK stars in both hemispheres (Blanco-Cuaresma et al. \cite{blanco}) that were analyzed with many different spectrum-synthesis codes and with data from different spectrographs. The analysis focused, among other topics, on global stellar parameters like effective temperatures and gravities (Heiter et al. \cite{heiter}), metallicity (Jofr\'e et al. \cite{jof1}) and chemical abundances (Jofr\'e et al. \cite{jof2}), and on some stars in detail (e.g., Creevey et al. \cite{creevey} on HD\,140283). A recent addition to low-metallicity stars was presented by Hawkins et al. (\cite{haw:jof}).

One of the major conclusions from this program is that even for main sequence stars the spectroscopically determined quantities may differ significantly with respect to the fundamental value obtained from direct radius and flux measurements. But even within the spectroscopic analyzes the global parameters $T_{\rm eff}$, $\log g$, and metallicity, along with micro- and macroturbulence and rotational line broadening, differ uncomfortably large (see, for example the discussion in Heiter et al. \cite{heiter}). Among the least cooperative stars appear to be the giants and metal-poor dwarfs. Comparisons indicate, for example that for solar-metallicity giants with $T_{\rm eff}\approx 5000$\,K and $\log g\approx 2.8$, the spectroscopic approach can lead to overestimated gravities. Intrinsic line-profile variability may become the limit for precision, in particular due to (non-detected) p-mode oscillations for the K giants and rotationally-modulated stellar activity (spots, faculae etc.) for all of the cool stars. Besides intrinsic stellar variability, one extra source of uncertainty is the remaining blending situation with weak molecular lines due to finite spectral resolution.

The data in the {\sl Gaia} benchmark spectral library provided by Blanco-Cuaresma et al. (\cite{blanco}) were originally homogenized to the highest minimum spectral resolution of the different spectrographs involved, and corresponded in practice to $R=\lambda/\Delta\lambda=70,000$. The vast majority of the spectra for the benchmarks are from HARPS (Mayor et al. \cite{harps}) with an original resolution of 115,000. NARVAL spectra (Auri\'ere \cite{narval}) have an average resolution of 65,000 (Paletou et al. \cite{pal:boe}), and the UVES data (Dekker et al. \cite{uves}) have 78,000--115,000. Peak signal-to-noise ratios (S/N) of between 150 to 824 were achieved. Higher S/N was obtained for eight bright stars in the UVES-POP library (Bagnulo et al. \cite{pop}) but with the lower spectral resolution of $\approx$80,000. Yet, if some of the most decisive astrophysical processes for stellar variability, that is, stellar convection, is to be detected and characterized in spectra, it requires much higher spectral resolution and even higher signal-to-noise ratio, and thus even larger telescopes. Most of the stars analyzed in work previous to the {\sl Gaia}-ESO survey had even lower resolution than the minimum resolution for the benchmark stars, for example the bright M-giant $\gamma$~Sge with 50--60,000, which is still the typical spectral resolution for a fixed-format \'echelle spectrograph on a smaller telescope.

\begin{table*}
\caption{Stellar sample of deep PEPSI spectra.} \label{T1}
\begin{tabular}{lllllllll}
\hline \noalign{\smallskip}
Star & M-K   &  \multicolumn{6}{c}{Combined S/N$^a$} & $N$ spectra$^b$  \\
     & class &  I/404 & II/450 & III/508 & IV/584 & V/685 & VI/825 &  \\
\noalign{\smallskip}\hline \noalign{\smallskip}
\emph{Giants} & & & & & & & & \\
\noalign{\smallskip}
\object{32 Gem}          & A9 III  &100 &260 &350 &470 &480 &420 & 222 222  \\
\object{HD 140283}       & F3 IV   &110 &210 &300 &460 &480 &660 & 222 224  \\
\object{HD 122563}       & F8 IV   &100 &240 &400 &770 &740 &750 & 222 224  \\
\object{$\eta$ Boo}      & G0 IV   &610 &980 &1400 &1700 &2180 &1660 & 233 36(10) \\
\object{$\zeta$ Her}     & G0 IV   &730 &240 &570 &640 &1470 &2040 & 286 56(14)  \\
\object{$\delta$ CrB}    & G3.5 III&210 &360 &500 &740 &790 &1050 & 122 223 \\
\object{$\mu$ Her}       & G5 IV   &520 &870 &990 &1500 &1630 &1730 & 28(13) 8(10)(13) \\
\object{$\beta$ Boo}     & G8 III  &500 &710 &1020 &1310 &1710 &1800 & 3(10)(10) (10)(10)(14)  \\
\object{$\epsilon$ Vir}  & G8 III  &330 &130 &220 &450 &730 &1970 & 363 33(11) \\
\object{$\beta$ Gem}     & K0 IIIb &400 &630 &1310 &1850 &2480 &2750 & 693 83(11)  \\
\object{HD 107328}       & K0 IIIb &150 &430 &630 &1110 &1240 &1700 & 343 436  \\
\object{$\alpha$ UMa}    & K0 III  &330 &180 &290 &640 &900 &1600 & 363 339 \\
\object{$\alpha$ Ari}    & K1 IIIb &680 &530 &640 &1630 &1250 &1400 & 366 866 \\
\object{$\alpha$ Boo}    & K1.5 III&500 &1500 &1990 &3000 &5780 &4330 & 6(14)4 7(46)(17) \\
\object{7 Psc}           & K2 III  &170 &470 &840 &1030 &1470 &1550 & 233 334 \\
\object{$\mu$ Leo}       & K2 III  &260 &430 &750 &940 &1400 &1850 & 234 34(16) \\
\object{$\gamma$ Aql}    & K3 II   &180 &500 &840 &1000 &1460 &1540 & 2(11)5 56(11) \\
\object{$\beta$ UMi}     & K4 III  &200 &270 &510 &750 &1140 &2100 & 433 33(12) \\
\object{$\alpha$ Tau}    & K5 III  &550 &1280 &1920 &2680 &3100 &3480 & 3(11)8 (12)8(10) \\
\object{$\mu$ UMa}       & M0 III  &190 &240 &520 &910 &1260 &2200 & 396 76(15) \\
\object{$\gamma$ Sge}    & M0 III  &240 &100 &380 &670 &1340 &2710 & 2(13)9 9(10)(17) \\
\object{$\alpha$ Cet}    & M1.5 IIIa&300 &280 &820 &1830 &1680 &1170 & 326 962 \\
\noalign{\smallskip}
\emph{Dwarfs} & & & & & & & & \\
\noalign{\smallskip}
\object{$\alpha$ CMa}    & A1 V    &800 &590 &720 &740 &620 &420 & 111 111 \\
\object{HD 84937}        & F2 V    &20 &110 &170 &230 &250 &200 & 352 523  \\
\object{$\sigma$ Boo}    & F4 V    &490 &190 &370 &420 &860 &1490 & 4(10)5 45(18) \\
\object{HD 49933}        & F5 V-IV &220 &390 &670 &740 &920 &620 & 126 262 \\
\object{$\alpha$ CMi}    & F5 V-IV &500 &410 &610 &1040 &1550 &1600 & 5(12)6 66(14) \\
\object{$\theta$ UMa}    & F7 V    &550 &240 &430 &660 &1110 &1670 & 4(14)8 88(20) \\
\object{$\beta$ Vir}     & F9 V    &250 &90 &150 &330 &440 &1220 & 273 43(10) \\
\object{HD 22879}        & F9 V    &180 &290 &260 &540 &380 &780 & 121 211 \\
\object{HD 189333}       & F9 V    &25 &130 &225 &170 &290 &270 & 111 111 \\
\object{HD 159222}       & G1 V    &170 &340 &450 &580 &710 &800 & 232 323 \\
\object{16 Cyg A}        & G1.5 V  &100 &250 &400 &530 &650 &600 & 122 223 \\
\object{HD 101364}       & G2 V    &40 &180 &300 &350 &470 &410 & 336 363 \\
\object{HD 82943}        & G2 V    &35 &180 &260 &340 &390 &330 & 322 223 \\
\object{18 Sco}          & G2 V    &240 &310 &590 &1040 &1300 &850 & 433 773 \\
\object{51 Peg}          & G2.5 V  &330 &710 &980 &1120 &1410 &1360 & 344 445 \\
\object{16 Cyg B}        & G3 V    &180 &340 &430 &700 &1110 &810 & 222 263 \\
\object{70 Vir}          & G4 V-IV &250 &450 &700 &910 &1160 &1340 & 222 224 \\
\object{$\mu$ Cas}       & G5 V    &220 &400 &670 &760 &980 &1000 & 122 222 \\
\object{HD 103095}       & G8 V    &100 &270 &450 &670 &700 &740 & 233 433 \\
\object{$\tau$ Cet}      & G8.5 V  &420 &820 &1200 &1650 &1910 &1570 & 566 896 \\
\object{$\epsilon$ Eri}  & K2 V    &410 &940 &1350 &1800 &2100 &1950 & 156 565 \\
\object{HD 192263}       & K2 V    &70 &200 &310 &400 &590 &530 & 122 222 \\
\object{HD 128311}       & K3 V    &20 &120 &190 &290 &350 &430 & 322 223 \\
\object{HD 82106}        & K3 V    &10 &75 &250 &240 &480 &360 & 213 132 \\
\object{61 Cyg A}        & K5 V    &105 &490 &690 &990 &1350 &1080 & 124 354 \\
\object{61 Cyg B}        & K7 V    &70 &200 &400 &540 &850 &980 & 122 224 \\
\noalign{\smallskip}\hline
\end{tabular}
\tablefoot{$^a$S/N per pixel at the continuum is given for all cross dispersers (CD, roman number) at the respective central wavelengths. For example, for CD\,I the central wavelength is 404\,nm. $^b$ $N$ denotes the number of individual exposures per cross disperser, for example ``222 222'' means two exposures in all six cross dispersers (a parenthesis is used if the number of spectra reached two digits). Note that if the LBT is used, then one exposure means two simultaneous individual spectra (one from each telescope). }
\end{table*}

\begin{figure*}
\includegraphics[angle=0,width=\textwidth,clip]{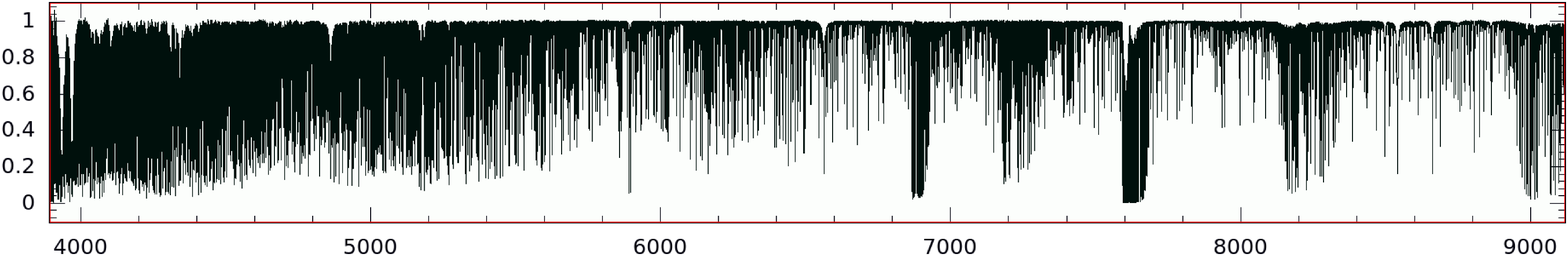}
\includegraphics[angle=0,width=\textwidth,clip]{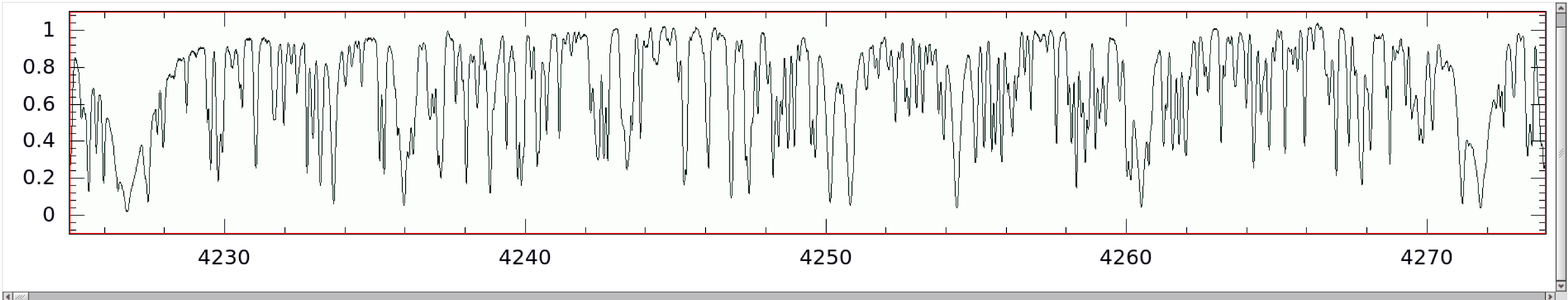}
\includegraphics[angle=0,width=\textwidth,clip]{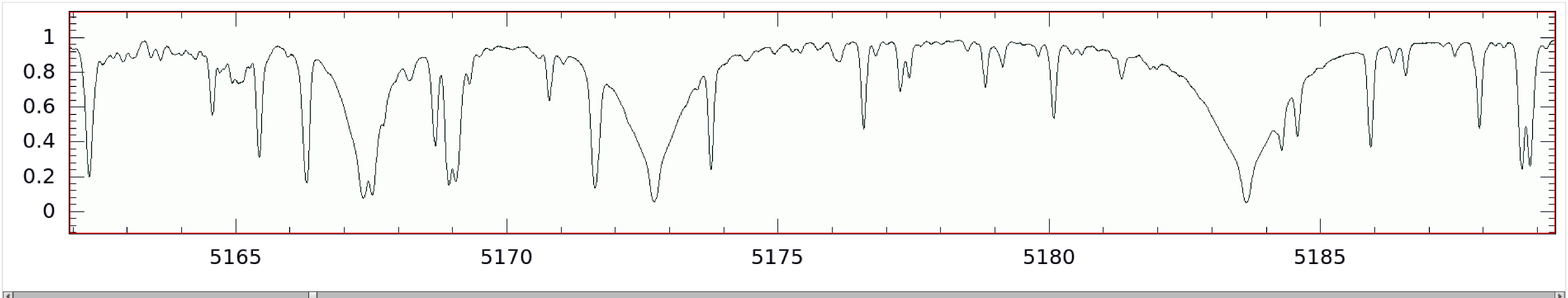}
\includegraphics[angle=0,width=\textwidth,clip]{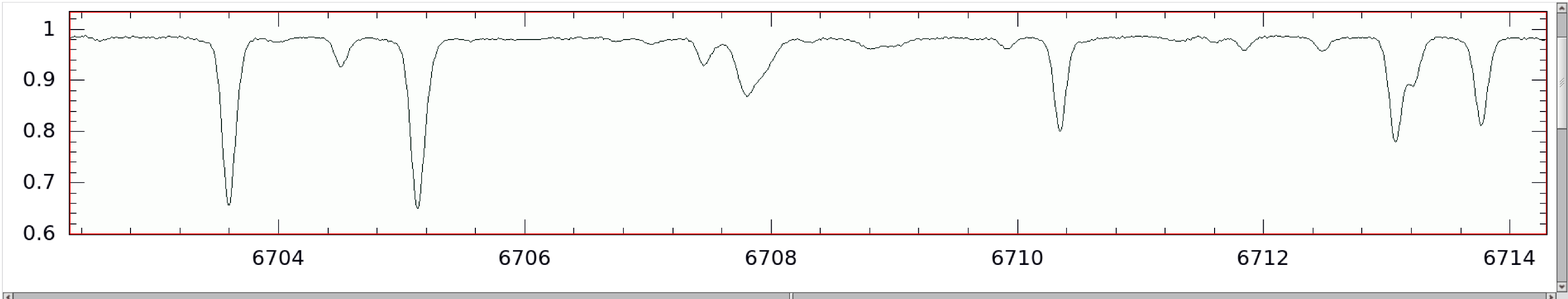}
\includegraphics[angle=0,width=\textwidth,clip]{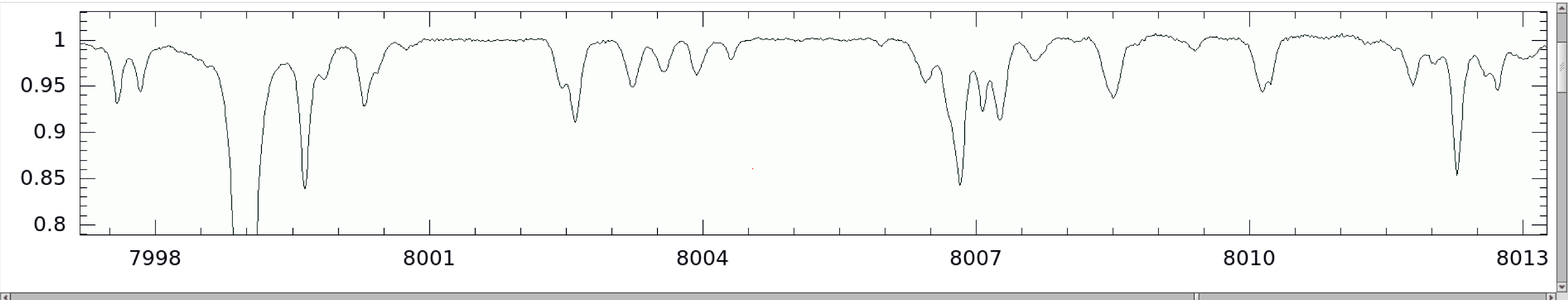}
\caption{Example spectrum of 70\,Vir, spectral type G4\,V-IV (x-axis is wavelength in Angstrom, y-axis is relative intensity). \emph{Top:} Full 5290-\AA\ wavelength coverage. The other panels show magnifications of four selected wavelength regions. \emph{Top down:} numerous Fe\,{\sc i} lines around 4250\,\AA , the Mg\,{\sc i} triplet at 5175\,\AA , the Li\,{\sc i} 6708-\AA\ region, and the 8004-\AA\ line region commonly used for $^{12}$C\,/\,$^{13}$C determination.}
 \label{F1}
\end{figure*}

We have just put into operation the optical high-resolution \'echelle spectrograph PEPSI (Potsdam Echelle Polarimetric and Spectroscopic Instrument) at the currently largest aperture optical telescope in the world; the effective 11.8m LBT (Large Binocular Telescope; Hill et al. \cite{lbt}). PEPSI provides a spectral resolution of up to 270,000 for the wavelength range 383--912\,nm and can alternatively be fed by the nearby 1.8m Vatican Advanced Technology Telescope (VATT) or by a small solar disk integration (SDI) telescope. The instrument is described in detail in Strassmeier et al. (\cite{pepsi}) and we refer to this paper for technical and operational details. Solar flux spectra with PEPSI were presented and analyzed in paper~I (Strassmeier et al. \cite{I.Sun}), where we also described the data reduction in more detail. A further application is presented in paper~III to Kepler-444 (Mack et al.~\cite{mack}). 

The current paper presents first PEPSI spectra for 48 bright stars including those of the original 34 \emph{Gaia} benchmark stars that were observable from the northern LBT site.  Data were taken during several instrument commissioning runs throughout 2015 until March 2017. Peak S/N of several thousand in the red wavelength regions were achieved with average-combining individual exposures. All these spectra are extremely rich in detail and therefore are made public for further data mining and analysis. Science-ready FITS files of the average-combined spectra are available for download\footnote{https://pepsi.aip.de} and raw data may be requested. Section~\ref{S2} describes the observations and the spectrograph while Sect.~\ref{S3} briefly reiterates the data reduction. Section~\ref{S4} gives a first quality assessment of the data product.  The stellar sample and its science cases with a few additions to the original {\sl Gaia} benchmark stars are discussed in Sect.~\ref{S5} and Sect.~\ref{S6}, respectively. Section~\ref{S7} is a summary.

\section{Observations}\label{S2}

High-resolution spectra with $R=\lambda/\Delta\lambda$ between 200,000 and up to 270,000 were obtained with PEPSI (Strassmeier et al. \cite{pepsi}) at the 2$\times$8.4\,m Large Binocular Telescope (Hill et al. \cite{lbt}). PEPSI is a fiber-fed white-pupil \'echelle spectrograph with two arms (blue and red optimized) covering the wavelength range 383--912\,nm with six different cross dispersers (CD). The instrument is stabilized in a pressure and thermally controlled chamber and is fed by three pairs of octagonal fibers per LBT unit telescope (named SX and DX). The different core diameters of the fibers and their respective image slicers set the three different resolutions of the spectrograph (43,000, 120,000 and 250,000). In use for this paper was image slicer block \#2. Its 250,000-mode, which was used for the spectra in this paper, is made possible with the seven-slice image slicers and a 100-$\mu$m fiber with a projected sky aperture of 0.74\arcsec , comparable to the median seeing of the LBT site. Its resolution element is sampled with two pixels. One resolution element corresponds to 1.2\,\kms\ for the resolving power of 250,000. Two 10.3k$\times$10.3k STA1600LN CCDs with 9-$\mu$m pixels record a total of 92 \'echelle orders in six wavelength settings. The dispersion changes from 8~m\AA/pixel at 400\,nm to 18~m\AA/pixel at 900\,nm and results in a $\approx$430,000 pixel long spectrum with unequally spaced step size. An example spectrum is shown in Fig.~\ref{F1}.

The six wavelength settings are defined by the six cross dispersers (CD\,I to VI), three per arm and two always simultaneously. It takes three exposures to cover the entire wavelength range 383-912\,nm. Several of the spectra extend out to 914\,nm because we repositioned cross disperser VI during commissioning and then gained another \'echelle order. Any of the three resolution modes can be used either with sky fibers for simultaneous sky and target exposures or with light from a stabilized Fabry-P\'erot etalon (FPE) for simultaneous fringe and target exposures for precise radial velocities (RV). However, the short integrations for the targets in this paper made the use of the sky fibers obsolete. Therefore, all spectra in this paper were taken in target+target mode, i.e., without simultaneous sky or FPE exposures, leaving two simultaneous target spectra on the CCD (one from each of the 8.4\,m LBT unit telescopes). We always refer to the combined target+target spectrum when referring to ``an LBT spectrum'', e.g.for example in the number of spectra in the tables. The SDI telescope makes use of PEPSI during day time and also consists of two spectra per exposure. Paper~I in this series dealt with these spectra. A 450-m fiber feed (Sablowski et al. \cite{vatt}) from the 1.8m VATT was used for very bright stars on several occasions. We note that the VATT spectra consist of just a single target spectrum on the CCD because it is only a single telescope.

Table~\ref{AT1} in the appendix summarizes the detailed observing log and identifies the individual spectra. Observations with the LBT were conducted during the epochs April~1-2 and 8-10, May~23-25, September~25 and 27, and November~19, in 2015; during June~2-4 and September~29-30 in 2016; and on March 3, 2017. Observations with the VATT were done during the epochs April~1-7, May~1-10, May~26-June~1, June~17-24, in 2015, and April~2-9 and May~24-June~2 in 2016. At around mid November 2015 the Blue CCD developed a bad amplifier (amp \#5) affecting a section of 5000$\times$1280 pixels on the CCD. It resulted in a drastically degraded quantum efficiency and practically in a loss of the three halves of the respective \'echelle orders that fell on this CCD section. This was repaired in early 2016 and the affected wavelength region was re-observed for some of the stars (but at a grossly different epoch). The spectrograph was in different states of alignment and focus and thus resulted in slightly different dependencies of spectral resolution versus wavelength. Also, acceptable barometric stability of the chamber was not achieved until summer 2015.

\section{Data reduction}\label{S3}

PEPSI data reduction was described in detail in paper~I. The Sun is the only target where we can compare to even higher quality data from Fourier Transform Spectrographs (e.g., Wallace et al. \cite{nso}, Reiners et al. \cite{iag}). The stellar PEPSI data in the present paper do not differ principally from the solar PEPSI data except for the S/N. This allows a direct comparison of solar flux spectra with stellar spectra because the solar light passes through exactly the same optical path as the stellar light. A full-wavelength PEPSI exposure in the $R$=250,000 mode is always merged from seven (image) slices per \'echelle order, 17 \'echelle orders per cross disperser, and two arms recorded with two different CCDs optimized for blue and red response.

The data were reduced with the Spectroscopic Data System for PEPSI (SDS4PEPSI) which is a generic software package written in C++ under a Linux environment and loosely based upon the 4A software (Ilyin \cite{4A}) developed for the SOFIN spectrograph at the Nordic Optical Telescope. It relies on adaptive selection of parameters by using statistical inference and robust estimators. The standard reduction steps include bias overscan detection and subtraction, scattered light surface extraction from the inter-order space and subtraction, definition of \'echelle orders, optimal extraction of spectral orders, wavelength calibration, and a self-consistent continuum fit to the full 2D image of extracted orders. Every target exposure is made with a simultaneous recording of the light from a Fabry-Perot etalon through the sky fibers. For the spectra in this paper, the Fabry-Perot wavelength solution was not implemented. Instead, exposures with a standard Thorium-Argon (Th-Ar) hollow-cathode lamp were used for the wavelength calibration. Its solution is based on about 5,000 spectral lines from all seven image slices and all spectral orders. A 3D Chebychev polynomial fit achieves an accuracy of about 3--5\,\ms\ in the image center with a rms of about 50\,\ms\ across the entire CCD. The Th-Ar calibration images are taken during daytime before the observing night and the wavelength scale zero point of the science images is thus relying on the stability of the PEPSI chamber. Wavelengths in this paper are given for air and radial velocities are reduced to the barycentric motion of the solar system. We note that the spectra in the present paper were reduced and extracted just like the Sun-as-a-star data in paper~I with the only difference that the continuum normalization was adjusted to the spectral classification.

In order to achieve the high S/N, we employ a ``super master'' flat-fielding procedure that is based on a combined flat-field image made from $\approx$2,000 individual images taken during daytime. Individual stellar spectra are first flat fielded, then extracted, and then average combined with a $\chi^2$ minimization procedure weighted with the inverse variance in each CCD pixel. For three taps on the STA1600LN CCDs surface, the pixel response is a function of the exposure level. This blemish is dubbed fixed-pattern noise and described in detail in paper~I. In order to achieve very high S/N, all raw images are corrected for nonlinearity of each CCD pixel versus its ADU level with the use of the super master flat field image. It comprises a polynomial fit of the CCD fixed-pattern response function in 70 de-focused flat field images at different exposure levels. Each of such an image is again the sum of 70 individual exposures made in order to minimize the photon noise. The fixed-pattern response function is the ratio of the flat field image and its 2D smoothed spline fit which removes all signatures of the de-focused spectral orders. The accuracy of this correction allows reaching a S/N in the combined spectra of as high as 4,500. Because the pattern shows a spatial periodicity with a spacing of 38 pixels, which is wavelength dependent on the red CCD (but not on the blue CCD), it leaves residual ripples with a periodicity of 38 pixels in the spectrum. It is being removed to the best possible level but the residual pixel-to-pixel non uniformity after the super master flat division currently limits the effective S/N from the affected CCD taps to $\approx$1,300:1. Other regions are unaffected.

\begin{figure*}
{\bf a.} \hspace{58mm} {\bf b.} \hspace{58mm} {\bf c.} \\
\includegraphics[angle=0,width=58mm]{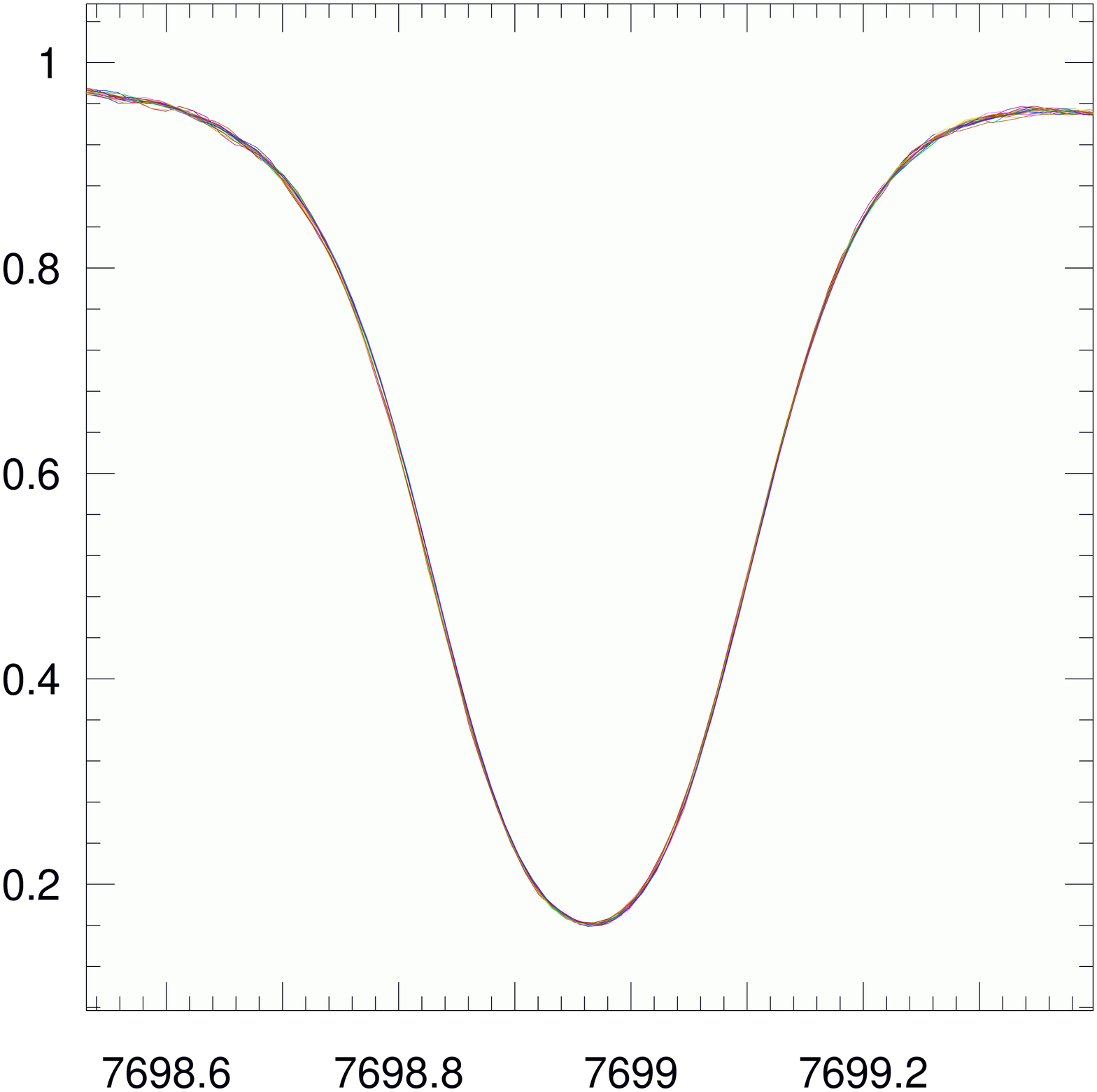}
\includegraphics[angle=0,width=58mm]{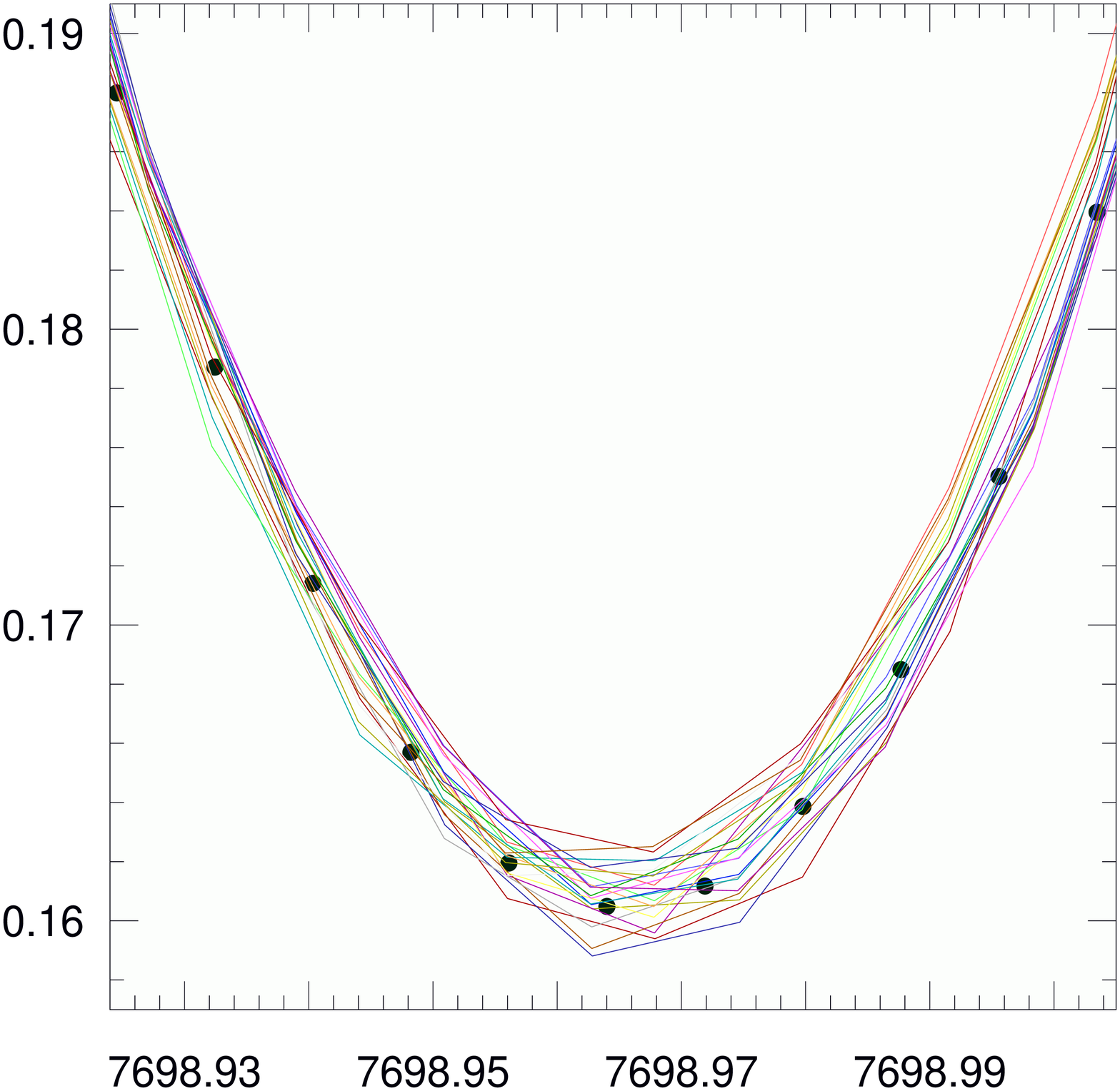}
\includegraphics[angle=0,width=61mm]{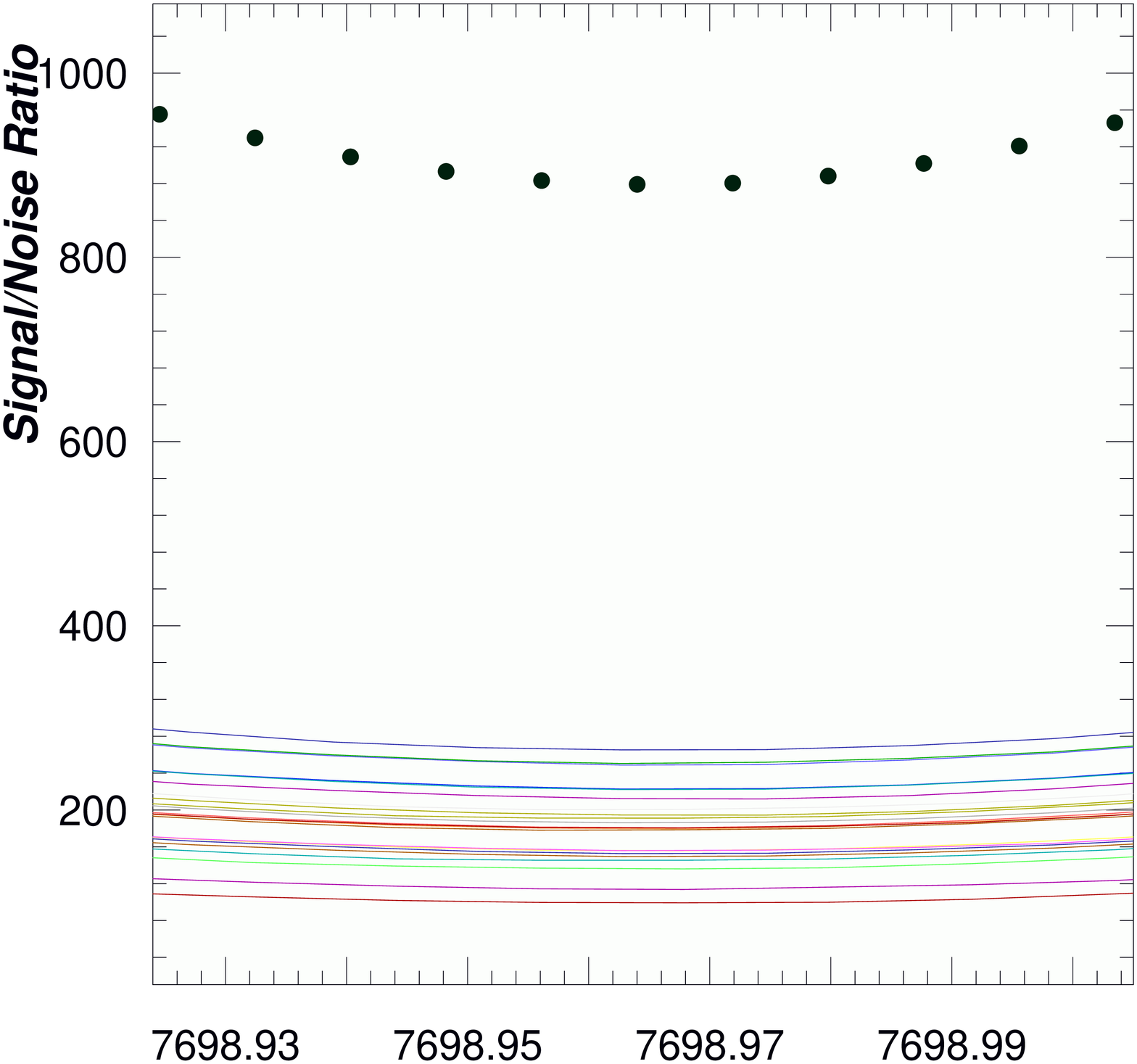}
\caption{Comparison of an average-combined spectrum and its individual exposures for $\beta$\,UMi (K4III). The x-axes are wavelengths in \AA . The 22 individual exposures are shown as lines, the average combined ``deep spectrum'' as dots. \emph{a.} A 0.8-\AA\ section showing the 22 exposures of the K\,{\sc i} 7699\,\AA\ line profile. The y-axis is relative intensity. No differences can be seen at this plot scale. \emph{b.} A zoom into a 0.08-\AA\ subsection of the line core. The spacing of the dots represent the CCD pixel dispersion. \emph{c.} S/N per pixel in the K\,{\sc i} line core for the same spectral window as in \emph{b}. }
 \label{F-bumi}
\end{figure*}

\section{Data product}\label{S4}

\subsection{Building deep spectra}

Deep spectra in this paper are build by average-combining individual spectra, usually taken back-to-back within a single night.  For some targets individual spectra are from different nights or even different runs. The observing log is given in Table~\ref{AT1}. Any spectrum co-addition is prone to variations of the entire spectrograph system, including errors from the seeing-dependent fiber injection or the temperature-sensitive CCD response or the optical-path differences due to the LBT being two telescopes and/or when using the VATT as the light feed.

When two spectra from different LBT sides are combined, the spectra are re-sampled into the wavelength grid of the first spectrum, then they are averaged with their corresponding weights (inverse variance in each pixel). First, the optimally extracted spectra from every slice of the image slicer are combined into a single spectrum for each spectral order. Prior to its averaging, all spectra are re-sampled into the common wavelength grid of the middle slice. The wavelength in each pixel is defined by the Th-Ar wavelength solution and remains unequally sampled. Then, the two average spectra from the two sides of the LBT telescope are combined into a single spectrum with the weighted average to the common wavelength grid of the first spectrum. One spectrum is re-sampled to the other by means of spline interpolation.

Because the spectra are normalized to a pseudo-continuum $I/I_c$, there is no need to re-scale them in intensity and the final continuum level will be adherent to the one with the higher S/N. The very same procedure is applied for building the average-combined deep spectra. For more details we refer to paper\,I.

\subsection{Intrinsic spectrum variations}

A deep spectrum may be affected by intrinsic stellar changes, for example due to rotational modulation by spots and plages, non-radial surface oscillations, or Doppler wobbling due to unseen stellar companions or exoplanets. However, exposure times are usually rather short compared to the typical stellar variability timescales. When using the LBT the exposure times are typically just tens of seconds up to a few minutes but exposure times with the VATT and its 450-m fiber link can be up to 60~minutes. Nevertheless, even after adding the CCD read-out overhead of 90\,s and the co-adding of, say, typically four spectra, the total time on target still sums up to a relatively short period in time, typically just tens of minutes for the LBT data and up to a maximum of three hours for the VATT targets. Any intrinsic changes with a timescale significantly longer than this we can safely dub non critical. This is certainly the case for rotational modulation and the revolution of orbital companions where we expect periods in excess of many days up to thousands of days. An example is $\alpha$~Tau, which shows RV variations of up to 140\,\ms\ with a period of 629\,d due to a super-Jupiter companion as well as rotational modulation with a period of 520\,d (Hatzes et al. \cite{hatzes15}).

Solar-like non-radial oscillations act on much shorter time scales than rotation and were discovered in several of our targets, for example for $\eta$~Boo (Kjeldsen et al. \cite{kjeldsen}). Frandsen et al. (\cite{frandsen}) detected oscillation periods of as short as 2–-5 hours in the giant $\xi$~Hya, another target in this paper. Just recently, Grundahl et al.~(\cite{grund}) identified 49 pulsation modes in the G5 subgiant $\mu$\,Her, originally discovered by Bonanno et al. (\cite{alfio}) as an excess of power at 1.2\,mHz. There is also evidence that sub-\ms\ RV oscillations of as short as 50 minutes were found in HARPS spectra of the active subgiant HR\,1362 (Dall et al. \cite{dall}) (not part of the target list in this paper). While the solar 5-min oscillation produces only a disk-averaged full RV amplitude of order 0.5\,\ms\ (see paper~I, but also Christensen-Dalsgaard \& Frandsen~\cite{chr:fra}, Probst et al.~\cite{probst}), this amplitude could amount to tens of \ms\ for K and M giants. For example, Kim et al. (\cite{kim}) found $\alpha$~Ari to be a pulsating star with a period of 0.84\,d and a RV amplitude of 20~\ms\ while Hatzes et al. (\cite{hatzes12}) found for $\beta$~Gem up to 17 pulsation periods with individual RV amplitudes of up to 10\,\ms . A mix of rotational and orbital, and pulsation periods of 230\,d and 471\,d, respectively, were found for the single-lined M0 giant $\mu$\,UMa (Lee et al. \cite{lee}).

In case there are line-profile and RV changes in our targets due to radial or non-radial oscillations, we would then obtain an intrinsically averaged spectrum. For one of the late-type giant targets, $\beta$~UMi (K4III), Fig.~\ref{F-bumi} shows a visual comparison of all 22 individual spectra (from 11 consecutive back-to-back exposures in binocular mode) with its combined spectrum. Only a single spectral line is shown; K\,{\sc i} 7699\,\AA\ ($\log gf = -0.176$, $\chi$= 1.6\,eV). Panels a and b are for two levels of magnified details, while the right panel c shows the S/N per pixel just for the line core. Note that the zoom factor in the plot in Fig.~\ref{F-bumi}b is considerable; all there is shown is a 0.08-\AA\ region centered at the K\,{\sc i}-7699 line core. The width of the entire plot is approximately one spectral resolution element of UVES or HARPS. We chose this line because of its absorption components from the local interstellar medium (LISM) which, if present, may be used as a reference with respect to the stellar photospheric contribution. No such LISM lines are seen in $\beta$\,UMi. Not unexpected, the largest spread of intensity is found with a 0.3\%\ peak-to-valley (p-v) after 15\,min on target (rms of 0.1\%), sampled with 11 consecutive 15-s exposures with both LBT telescopes. For the line core, this spread is expected for the data quality of a single exposure of typical S/N of $\approx$200 in the K\,{\sc i}-line core. The average-combined spectrum has a S/N of 2,250 in the continuum on both sides of the line and still around 900 in the line core itself (Fig.~\ref{F-bumi}c). Therefore, a p-v spread of 0.3\%\ near the continuum is a significant deviation with respect to the photon noise and is likely due to the combined effects of remaining systematic errors (mostly continuum setting) and yet unverified intrinsic stellar variations, presumably  non-radial pulsations.

\subsection{Signal-to-noise ratio}\label{snr}

Table~\ref{T1} lists the S/N for the deep spectra. S/N is always given for the continuum and per pixel at the mid wavelength of each cross disperser. The variance of each CCD pixel in the raw image is originally calculated after bias subtraction given the measured CCD gain factor for each of the 16 amplifiers of the STA1600LN CCD. The original variance is then propagated at every data reduction step and its final value is used to determine the S/N in every wavelength pixel of the final continuum normalized spectrum. The number of individual spectra employed for a deep-spectrum varies from star to star and is given in the last column of Table~\ref{T1}. It may differ from Table~\ref{AT1} because not always all existing spectra were included for the combined average spectrum. In case of bad seeing when spectra had significantly lower S/N, or when we had better spectra from different epochs, we simply did not use all of the available spectra but still list them in Table~\ref{AT1}. Several of our dwarf stars (HD\,84937, HD\,189333, HD\,101364, HD\,82943, HD\,128311, and HD\,82106) have S/N below 70:1 at the mid wavelength of CD-I at 400nm and are thus not recommended for a quantitative analysis at these wavelengths. Note that due to the relatively small fiber entrance aperture of 0.74\arcsec\ the S/N strongly depends on seeing. Blue-arm spectra that were taken only with the VATT and its 450\,m fiber generally have lesser S/N. The fibre transmission at 500\,nm is 30\% (see Strassmeier et al. \cite{pepsi}, their Fig.~41).

\begin{figure*}
{\bf a.}\\
\includegraphics[angle=0,width=\textwidth,clip]{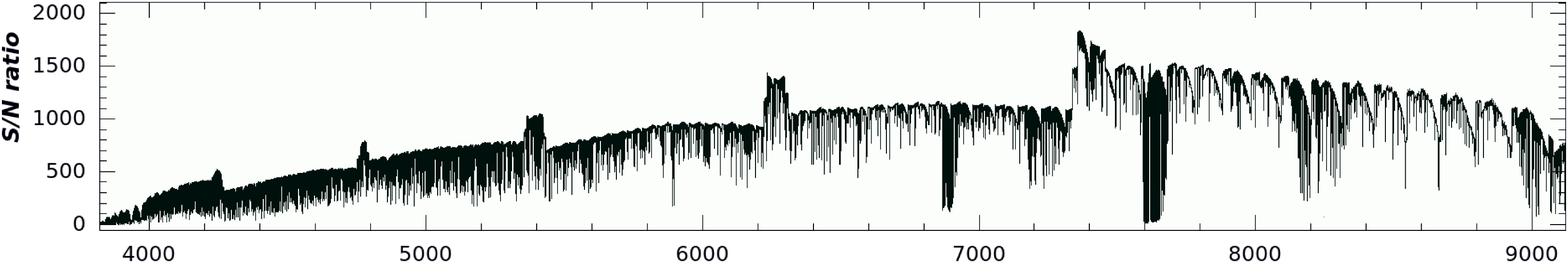}
{\bf b.}\\
\includegraphics[angle=0,width=\textwidth,clip]{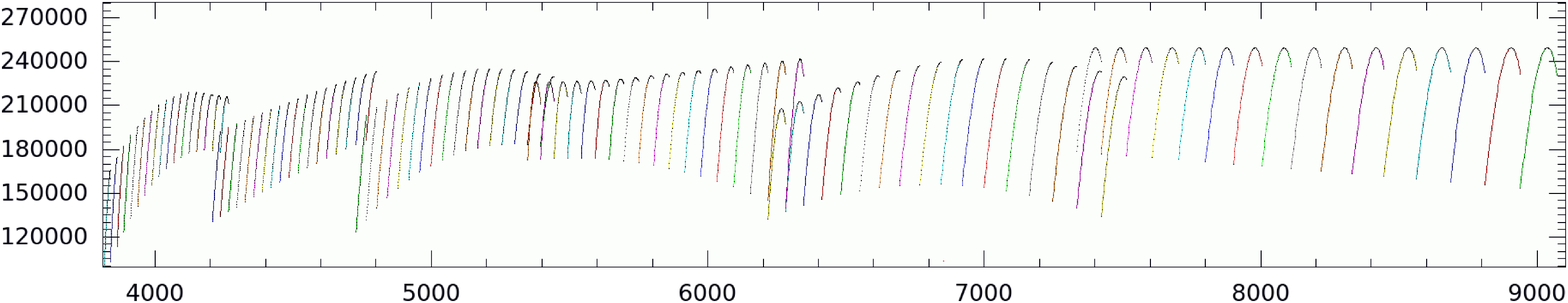}
\caption{Examples for the typical S/N and spectral resolution of the library stars in this paper. \emph{a.} Shown is the S/N of the deep spectrum of the G4 subgiant 70\,Vir. The larger S/N in the very red wavelengths is mostly due to the larger number of individual spectra. Note that the local peaks in S/N are due to the wavelength overlap of the cross dispersers, which effectively doubles the number of pixels there. \emph{b.} Shown is the spectral resolution for the focus achieved for the July 2016 run. }
 \label{F-snr}
\end{figure*}

Fig.~\ref{F-snr}a is a plot of a typical distribution of S/N across wavelength for one of the targets. Note that wavelength regions at the edges of spectral orders repeat in adjacent orders. Because the order merging procedure in our SDS4PEPSI pipeline is a weighted average of the overlapping wavelength regions, the S/N for these strips of wavelength is a weighted average from two different orders and thus formally higher than the rest, but not necessarily more accurate due to continuum-setting uncertainties.

\subsection{Spectral resolution}

As in any \'echelle spectrograph, the resolution changes along and across each spectral order due to the change of the optical image quality and the anamorphism of the grating. The optical design of PEPSI includes a small $0.7^\circ$ off-plane angle of the incident beam, which results in a small tilt of $4^\circ$ of the spectral lines formed by the image slicers. This tilt is compensated by a fixed counter-rotation of the entire image slicer. The residual tilt of the slices that are further away from the optical axis is of the order of $1^\circ$, which degrades the resolving power only by $\approx$2\%. The main contributor to the resolving power variations in PEPSI is the change of the optimal focus position across the large area of the two 10k CCDs.

The nominal spectrograph resolving power versus wavelength was shown in our paper~I. The true resolution varied from run to run, in particular for the data taken during commissioning because the line FWHM and, thus, the resolution is computed from the Th-Ar lines at the focus positions achieved at that time. Specifically, we calculated the resolving power for every Th-Ar emission line from its FWHM in pixels and the dispersion at the line position derived from the 3D dispersion polynomial. The best focus position is selected as the maximal median resolving power of all spectral orders versus focus position. An example for July 2016 is shown in Fig.~\ref{F-snr}b. Focus sequences for every observing run and for all cross-dispersers were made with the Th-Ar cathode as the light source. A FFT auto-correlation for a central region on the CCD with 3,000$\times$300 pixels is applied for focus determination, its FWHM is then converted to effective spectral resolution as a function of wavelength. Our spectra in this paper were all taken with a focus selected to suit all CDs at once. One could do a wavelength dependent focus, that is, refocus for every CD combination, but that was not done for the present data. Therefore, the average resolution for the stars in this paper falls mostly short of $\lambda/\Delta\lambda$ of 250,000 with a range from 180,000 near the blue cut-off to a peak of 270,000 near 700\,nm. The grand average from 383 to 912\,nm is approximately 220,000 or  1.36\,\kms . When modeling individual spectral regions, or line profiles, it is advisable to consult Fig.~\ref{F-snr}b or the FITS header for the true resolution at a particular wavelength. Resolution dispersion is approximately $\pm$30,000 for the entire 383 to 912\,nm range.

\subsection{Continuum}

Continuum-setting errors are among the largest uncertainties for high-resolution spectroscopy. It is particularly problematic for cool stars where line blanketing effectively removes any clean continuum signature. This is typical for blue wavelengths short of $\approx$450\,nm as well as those red wavelengths that are affected by the wide terrestrial O$_2$ absorption bands. It is generally the case for M stars due to increasing molecular bands with decreasing effective temperature. For solar-like stars in this paper, we referenced to the dry NSO FTS solar spectrum for continuum definition just like we did in paper~I for the Sun itself. For $\approx$K1 giants, we use as a reference the KPNO Arcturus atlas from Hinkle et al. (\cite{hinkle}). The final continuum is obtained from synthetic spectra. We tabulated synthetic spectra generated with the MARCS model atmospheres (Gustafsson et al. \cite{gus}), the VALD atomic line list (Kupka et al. \cite{vald}), the SPECTRUM code (Gray \& Corbally \cite{gra:cor}), and the atmospheric parameters from the literature (taken from Blanco-Cuaresma et al. \cite{blanco}). These synthetic spectra are first trimmed to the reference stars (Sun and Arcturus) and then employed to build a continuum free of spectral lines, which is then used for division for all targets of the atlas. We noticed a systematic opacity deficiency for giant-star spectra near our blue cut-off wavelength. Currently, we bypassed this by fitting the Arcturus atlas and deriving a fudge parameter that is then multiplied with the synthetic continuum.

There will always be residual continuum errors left in the data. For the red spectra these are generally caused by the increasing telluric contamination towards longer wavelengths. Although a small effect, it can cause systematic shifts at some wavelength regions of the order of 1\%. However, such small shifts may simply be compensated posteriori to the data reduction. More problematic is the line blanketing in the blue together with the lowered S/N of the blue spectra in general. It can lead to continuum errors of 10-20\%\ for the regions bluer than 397\,nm. Targets with very low S/N near the blue cutoff wavelength may appear with  line depths that are negative and peaks that are above continuum. When comparing these PEPSI spectra to synthetic spectra it is advisable to also refit the continuum level in question.

\begin{figure*}
{\bf a.}\\
\includegraphics[angle=0,width=\textwidth,clip]{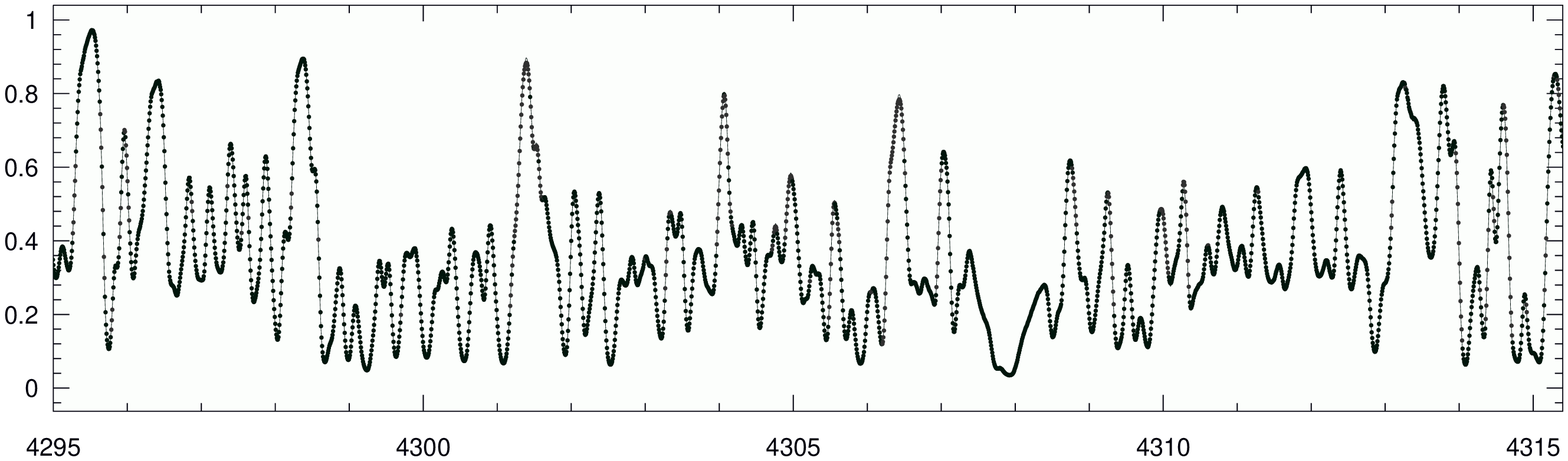}
{\bf b.}\\
\includegraphics[angle=0,width=\textwidth,clip]{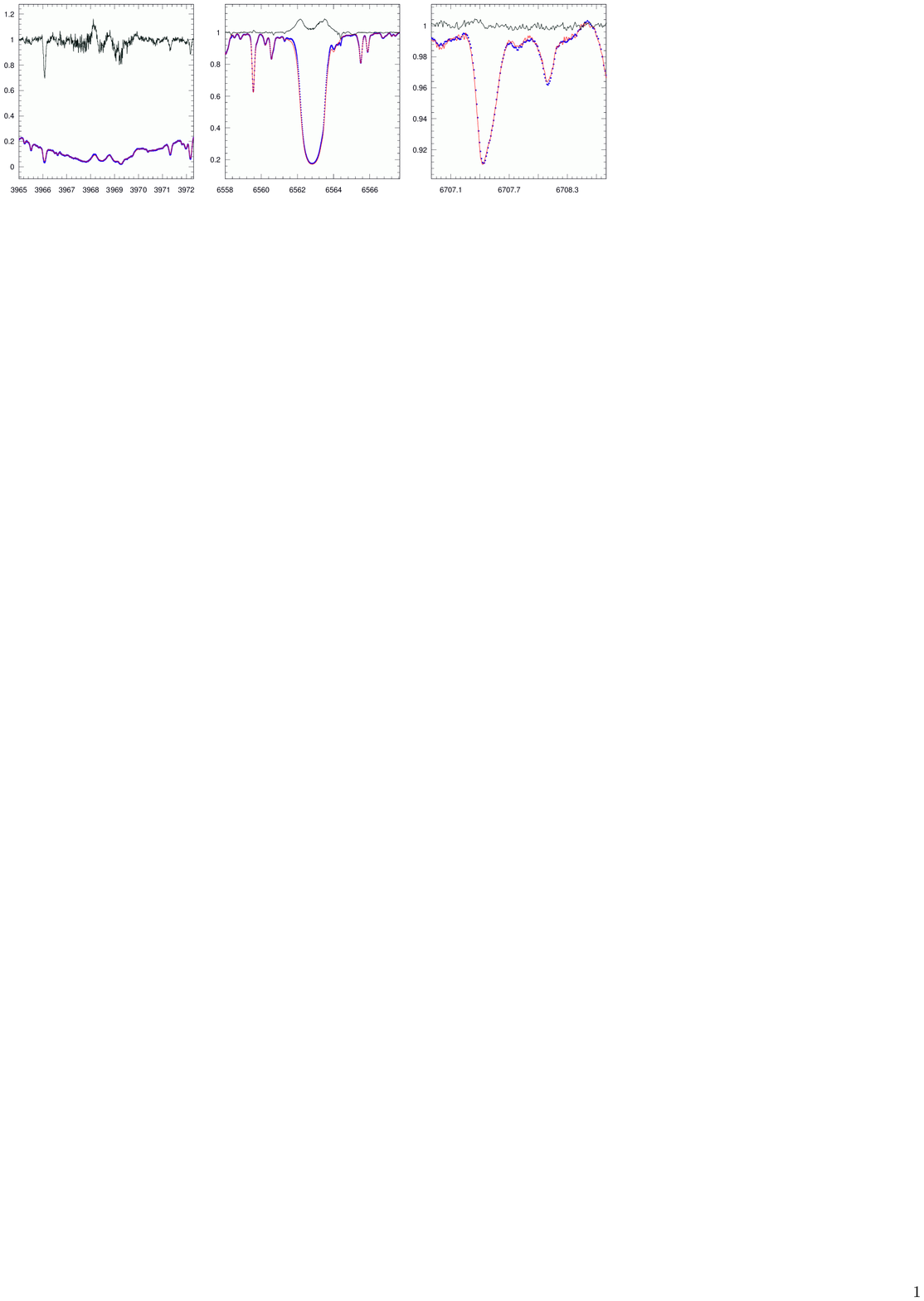}
\caption{Comparison of the deep PEPSI spectrum of Arcturus with the KPNO Arcturus atlas from Hinkle et al. (\cite{hinkle}). \emph{a.} Shown are the numerous CH lines centered around the Fe\,{\sc i}/Ca\,{\sc i} blend at 4308\,\AA\ that constitute the Fraunhofer $G$-band. The dots are the PEPSI spectrum and the line is the KPNO atlas. The match is nearly perfect. \emph{b.} Three wavelength regions where chromospheric activity may be detected. Each panel shows the ratio spectrum PEPSI:KPNO (line around unity), the KPNO spectrum as a line, and the PEPSI spectrum as dots. From left to right, the core of the \ion{Ca}{ii} H line, \Halpha , and the Li\,{\sc i} 6707.9-\AA\ line region. }
 \label{F-arct}
\end{figure*}

\subsection{RV zero point}

In paper~I, we have shown that our absolute RV zero point was within a rms of just 10\,\ms\ of the HARPS laser-comb calibrated solar atlas (Molaro et al. \cite{molaro}). This allows us to combine spectra taken over a long period of time, at least as long as the spectrograph chamber is not opened. Besides, there is a small systematic difference of on average 9\,\ms\ between the zero points of LBT SX and LBT DX spectra due to its separate light coupling. The shift between SX and DX is not a single constant but depends on the CD, ranging between 5 to 30\,\ms . The major contribution is because SX and DX have different foci, hence the stellar images can be offset from one another. We will investigate further whether the shift can be compensated with the simultaneous FPE reference source. For the present paper, any differential RV shifts were removed for each individual CD by means of a least-squares minimization with respect to the SX side. Larger than expected shifts occurred during commissioning because we realigned the CDs once in the blue arm and twice in the red arm, and the spectrum image had unavoidably moved on the detector. Despite that these data periods were treated with special attention, it created residual RV inhomogeneities of the order of the uncertainties themselves. Because we do not intend to provide RVs for the library stars, we shifted all spectra in this library to their respective barycentric RV value by subtracting the RV given in the header of the respective FITS file. This shall enable easier inter comparisons. The date given in the average combined FITS file is the average date resulting from the combination of the individual CDs, it has no physical meaning.

\subsection{Wavelength coverage}

The free spectral range for each CD has been graphically shown in our technical paper in Strassmeier et al. (\cite{pepsi}) as well as in paper~I, and we refer the reader to these papers for more details. Here we recap the wavelength coverage in nm for each cross disperser; CD-I 383.7--426.5, CD-II 426.5--480.0, CD-III 480.0--544.1, CD-IV 544.1--627.8, CD-V 627.8--741.9, and CD-VI 741.9--912. All spectra in this paper cover the full wavelength range 383--912\,nm. Because of the comparably low efficiency in the bluest cross disperser (CD-I), and the accordingly longer exposure times, its exposures had sometimes to be done on different nights for some targets. For a total of six dwarf-star targets, we lacked the telescope time to redo the CD-I exposure with an appropriate longer integration time (see Sect.~\ref{snr}). Their initial spectra provided only poor S/N for wavelengths shorter than or around Ca\,{\sc ii} H\&K, and we recommend not to use these wavelength sections for the six stars. For one of these target, HD\,82106, we did not include the spectrum shorter than 398\,nm because it affected the continuum setting for the remaining wavelengths of CD-I. For the other five stars, we decided to keep these wavelengths in the library for the sake of completeness. Note that the bluest wavelength regions are not accessible with the VATT because of the 450-m fibre link and its effectively 95\%\ absorption loss at 400\,nm. The stars affected can also be identified in Table~\ref{AT1} by their two-digit S/N for CD\,I/404nm.

\subsection{Blemishes}

Four stars observed in the time frame November-December 2015 have only a fractal wavelength coverage in the blue arm. On Nov.\,13,~2015, we discovered a degrading amplifier in the blue 10k~CCD that left its section on the CCD (approximately 5000$\times$1280 pixels) with an effective gain of 1.2 instead of 0.5. This in turn resulted in a non-linear behavior at our exposure levels and lowered the quantum efficiency, which both caused intensity jumps in the spectrum at the amplifier edges and thus confused the order tracing algorithm of the data reduction. It affects one half of altogether three \'echelle orders. The jumps were corrected and spectra extracted but with approximately half the S/N for the regions affected. The chip was repaired in early 2016 and the problem did not occur again (bonding at the CCD had gotten loose and introduced a variable ohmic resistance). The consequence of this is that we do not recommend to use the wavelength regions of this section of the blue CCD of the four stars (51~Peg, 7~Psc, $\alpha$~Ari, and $\alpha$~Cet), which were all taken during above time window. Table~\ref{T-blemish} lists the detailed wavelengths affected.

\begin{table}
\caption{Wavelength regions in \AA\ that are affected by a bad-amplifier problem with the Blue CCD for the epoch November-December 2015.} \label{T-blemish}
\begin{tabular}{llllll}
\hline \noalign{\smallskip}
\#$^a$ & CD-III & \# & CD-II & \# & CD-I \\
\noalign{\smallskip}\hline \noalign{\smallskip}
\noalign{\smallskip}
120      & 5063--5103 & 136 & 4484--4501 & 151 & 4025--4056 \\
119      & 5106--5145 & 135 & 4501--4534 & 150 & 4052--4083 \\
118      & 5150--5161 & 134 & 4534--4570 & 149 & 4080--4096 \\
\noalign{\smallskip}\hline
\end{tabular}
\tablefoot{$^a$\'Echelle order. Stars (CDs) affected are 51\,Peg~(I,II), 7\,Psc~(I,II,III), $\alpha$\,Ari~(II,III), and $\alpha$\,Cet~(II,III). }
\end{table}

Several of the spectra show a (time-variable) HeNe-laser emission line at 6329\,\AA . This line was picked-up accidentally as stray light from the LBT laser tracker system (which should have been turned off). Its FWHM is around 53\,m\AA\ and its shape appears rather asymmetric with an emission intensity of up to 1.4 relative to the (stellar) continuum. Spectra are usually not affected unless the laser line falls within a stellar line like, for example in the 61~Cyg~A spectra taken during the May 2015 run.

\subsection{A comparison with the KPNO Arcturus atlas}

Hinkle et al. (\cite{hinkle}) assembled an atlas spectrum of Arcturus from a large number of individual spectral segments obtained with the KPNO coud\'e feed telescope and spectrograph at Kitt Peak. It took 15 nights to complete all exposures (from Apr.~20 to June~13, 1999). Its spectral resolution ranges between 130,000 to 200,000 for the wavelength range 373 to 930\,nm and has been an outstanding reference for more than a decade. It succeeded the photographic Arcturus atlas of Griffin (\cite{griffin}).

We employ the KPNO atlas spectrum to find its continuum and use it for PEPSI division. Then, the PEPSI spectrum matches the KPNO spectrum almost perfectly. Fig.~\ref{F-arct}a shows a comparison of the KPNO atlas and the PEPSI spectrum for the Fraunhofer $G$-band. PEPSI's slightly higher spectral resolution shows up with slightly deeper lines, on average 0.5\%\ deeper than the KPNO lines (in particular in the red wavelengths). The wavelength distribution of the achieved S/N is rather inhomogeneous for the PEPSI spectrum because CD-III was employed for only four short VATT exposures while CD-VI had included nine LBT exposures (exposure time with the LBT was 3\,s, with the VATT 4\,min for similar S/N). Red ward of $\approx$400\,nm the photon noise is not noticeable anymore in the plots, peaking at 6,000:1 at 710\,nm, but is inferior than the KPNO atlas for the very blue parts shorter than $\approx$397\,nm, where values below 100:1 are reached.

The wavelength zero point of the KPNO and the PEPSI spectra differs by several hundred \ms . A cross correlation of each PEPSI spectrum with the appropriate part of the KPNO atlas suggests the following shifts PEPSI-minus-KPNO in \ms\ (CD-I 545, CD-II 631, CD-III 576, CD-IV 740, CD-V 783, CD-VI 871). Note that the higher differences for the red CDs are due to the increase of telluric lines in these wavelength regions which makes the cross correlation more prone to systematic errors. When comparing the two spectra these shifts were applied so that both spectra are in barycentric rest frame. The pixel-to-pixel ratio spectrum (PEPSI spectrum divided by the KPNO atlas) is in general dominated by the variable water content of Earth's atmosphere which creates numerous artifacts. Three wavelength sections containing the Ca\,{\sc ii} H line core, \Halpha, and Li\,{\sc i} 6708 are shown in Fig.~\ref{F-arct}b. The presence of emission in the cores of these lines is a simple diagnostic of magnetic activity in the chromospheres of late-type stars. As can be seen in Fig.~\ref{F-arct}b, the ratio spectrum shows weak differences in the Ca\,{\sc ii} H and \Halpha\ profiles at a few per cent level (10\%\ peak-to-valley in the H\&K line core and 8\%\ in \Halpha). Note that the time difference between the KPNO spectrum and the PEPSI spectrum is 16 years. We usually interpret such residual emission due to changing magnetic activity. While just a preliminary result, it supports the earlier claims of the existence of both \ion{Ca}{ii} H\&K variability (with a period of $\leq$14\,yrs) by Brown et al. (\cite{brown}) and the detection of a (very weak) longitudinal magnetic field by Hubrig et al. (\cite{hub}) and Sennhauser \& Berdyugina (\cite{sen:ber}).

\section{Notes on the stellar sample}\label{S5}

The sample in Table~\ref{T1} consists mostly of the \emph{Gaia} FGK benchmark stars (Blanco-Cuaresma et al. \cite{blanco}) that are accessible from the northern hemisphere. These are complemented by a few  well-known Morgan-Keenan spectroscopic standard stars from the list of Keenan \& McNeil (\cite{kee:mcn}) and partly available in the PASTEL database (Soubiran et al. \cite{pastel}). A few stars were added that we had kept as serendipitous references from the STELLA binary survey (Strassmeier et al. \cite{orbits}).

\subsection{Giants}

Chemical abundances and evolutionary indicators of giants are widely used as probes for the galactic chemical evolution but their precision and accuracy continue to remain a challenge (Gustafsson \cite{gus}, Luck \& Heiter \cite{luc:hei}, Luck \cite{luck}). We provide spectra for 22 red giant branch (RGB) stars and subgiants in the present library. These are mostly very bright and well-studied stars like Arcturus ($\alpha$\,Boo) or Procyon ($\alpha$\,CMi) but include a few less well-studied targets like 32\,Gem (=HD\,48843, HR\,2489) or 16\,Vir (=HD\,107328). Spectral M-K classifications are rather mature for these bright targets except maybe for 32\,Gem which had been listed as A9\,III by Morgan (\cite{morgan}), A7\,II by Cowley \& Crawford (\cite{cow:cra}), A9\,II-III by Cowley \& Fraquelli (\cite{cow:fra}) and, most recently, as A8\,II by Fekel (\cite{fcf}). Given the deep and structured Na\,D and K\,{\sc i} absorption profiles it must a distant and thus likely a bright-giant star. Another target, $\gamma$\,Aql (HD\,186791), has been consistently identified in the literature as a K3 bright giant of luminosity class II since the fifties of the last century (Morgan \& Roman \cite{mor:rom}), although Keenan \& Hynek (\cite{kee:hyn}) assigned a K3 supergiant classification based on infrared spectra. Three M giants comprise the very cool end of our classification sequence; two M0\,III stars ($\mu$\,UMa = HD\,89758 and $\gamma$\,Sge = HD\,189319) and one M1.5\,III ($\alpha$\,Cet = HD\,18884). Just recently, Lee et al. (\cite{lee}) reported secondary RV variations of the (single-lined) spectroscopic binary $\mu$\,UMa and concluded on an origin from pulsations and chromospheric activity. Planetary companions were found for the two K giants $\mu$\,Leo and $\beta$\,UMi (Lee et al. \cite{lee1}). Lebzelter et al. (\cite{leb}) presented a comparative spectroscopic analysis of two cool giants that are are also in the present paper; $\alpha$\,Tau (K5\,III) and $\alpha$\,Cet.

\begin{table}
\caption{Example results from PEPSI using ParSES.} \label{T-parses}
\begin{tabular}{llllll}
\hline \noalign{\smallskip}
         & $T_{\rm eff}$ & $\log g$ & [Fe/H] & $\xi_t$ & $v\sin i$ \\
         & (K)           & (cgs)    & (solar)& (\kms ) & (\kms ) \\
\noalign{\smallskip}\hline \noalign{\smallskip}
\multicolumn{6}{c}{Sun}\\ \noalign{\smallskip}
PEPSI  & 5730 & 4.44 & --0.04 & 1.2 & 0.0 \\
       & $\pm$50 & $\pm$0.15 & $\pm$0.15 & $\pm$0.2 & $\pm$1\\
\noalign{\smallskip}\hline \noalign{\smallskip}
\multicolumn{6}{c}{70 Vir}\\ \noalign{\smallskip}
PEPSI  & 5475 & 3.86 & --0.13 & 1.29 & 0.0 \\
       & $\pm$50 & $\pm$0.15 & $\pm$0.1 & $\pm$0.1 & $\pm$1\\
FOCES  & 5481 & 3.83  &--0.11  & \dots &  1.0 \\
       & $\pm$70 & $\pm$0.10 & $\pm$0.07 &  & $\pm$?\\
ELODIE & 5559 & 4.05 & --0.06 & 1.11 & 1.36 \\
       & $\pm$19 & $\pm$0.04 & $\pm$0.02 & $\pm$0.06 & $\pm$0.45\\
\noalign{\smallskip}\hline \noalign{\smallskip}
\multicolumn{6}{c}{$\alpha$ Tau}\\ \noalign{\smallskip}
PEPSI  & 3900 & 1.45 & --0.33 & 1.16 & 3.5 \\
       & $\pm$50 & $\pm$0.3 & $\pm$0.1 & $\pm$0.1 & $\pm$1.5\\
Benchmark &3927  & 1.11 &--0.37  &1.63 &5.0  \\
       & $\pm$40 & $\pm$0.15 & $\pm$0.17 & $\pm$0.30 & \dots\\
\noalign{\smallskip}\hline
\end{tabular}
\tablefoot{$\xi_t$ is the microturbulence; FOCES refers to Fuhrmann et al. (\cite{beso}). ELODIE refers to Jofr\'e et al. (\cite{jof1}). Benchmark refers to Heiter et al. (\cite{heiter}) for $T_{\rm eff}$ and $\log g$, and to Jofr\'e et al. (\cite{jof1}) for metallicity.}
\end{table}

\subsection{Dwarfs}

Spectra for 26 dwarfs are presented. Note that the benchmark star HD\,52265 (G0V) was observed only during a bad weather epoch with the VATT and therefore we do not include it into the PEPSI library but still keep its entry in the observing log in the appendix in Table~\ref{AT1}. During a search for the white dwarf companion of $\alpha$\,CMa (Sirius), we took a spectrum of its A1\,V primary, the brightest star in the sky, which we include in this paper for the sake of spectral mining. HD\,189333 = BD+38\,3839 ($V$=8\fm5), a star similar and close to the famous planet-transit star HD\,189733, was classified F5\,V from photographic spectra by as long ago as Nassau \& MacRae (\cite{nas:mac}), which is pretty much all there is known for this star. We took a single spectrum of it and make it available in the library. Another target, HD\,192263, has a checkered history as a planet host, see the full story in the introduction in the paper by Dragomir et al. (\cite{dra:kan}). Its planet was found, lost, and re-found in the course of a couple years. Our spectrum shows the star with strong Ca\,{\sc ii} H\&K emission and thus being chromospherically very active. Another planet-host star recently revisited in the literature (McArthur et al. \cite{upsAnd} and references therein) is HD\,128311=$\upsilon$\,And. Together with $\epsilon$\,Eri it is among the targets known to contain planets and debris disks.

\section{Representative science examples}\label{S6}

\subsection{Global stellar parameters}

Global stellar surface parameters like effective temperature, gravity, and metallicity are basics for our understanding of stars. Observed spectra are usually compared to synthetic spectra from model atmospheres, analog to what had been exercised for the benchmark stars of the {\sl Gaia}-ESO survey and many other such attempts in the literature (e.g., Paletou et al. \cite{pal:boe}). The main advantage to do so again for benchmark stars is the homogeneity of the data in this paper and their significantly higher spectral resolution. There is no need to homogenize the data by artificially broadening the spectra to match the resolution of the lowest contributor. For a first trial, we employ our spectrum synthesis code ParSES to two selected deep spectra; 70~Vir (G4V-IV) and $\alpha$~Tau (K5III). ParSES is based on the synthetic spectrum fitting procedure of Allende-Prieto et al. (\cite{all}) and described in detail in Allende-Prieto (\cite{all04}) and Jovanovic et al. (\cite{parses}).

Model atmospheres were taken from MARCS (Gustafsson et al. \cite{gus}). Synthetic spectra are pre-tabulated with metallicities between --2.5\,dex and +0.5\,dex in steps of 0.5\,dex, logarithmic gravities between 1.5 and 5 in steps of 0.5, and temperatures between 3500\,K and 7250\,K in steps of 250\,K for a wavelength range of 380-–920\,nm. This grid is then used to compare with selected wavelength regions. For this paper, we used the wavelength range of only one of the six cross dispersers (CD\,IV 544.1--627.8\,nm). We adopted the {\sl Gaia}-ESO clean line list (Jofre et al. \cite{jof1}) with various mask widths around the line cores between $\pm$0.05 to $\pm$0.25\,\AA . Table~\ref{T-parses} summarizes the best-fit results. We emphasize that these values are preliminary and meant for demonstration.

We first applied ParSES to the PEPSI solar spectrum, same wavelength range and line list as for the other two example stars. It reproduces the expected basic solar parameters very well (Table~\ref{T-parses}). Its parametric errors based on the $\chi^2$ fit are likely not representative for the other stars because we adopted the NSO FTS continuum for rectification.

70\,Vir is more a subgiant than a dwarf. A $\log g$ of 3.89 was given by Fuhrmann et al. (\cite{beso}) based on the Hipparcos distance from which the iron ionization equilibrium temperature of 5531\,K results.  A previous FOCES analysis by Bernkopf et al. (\cite{bern}) gave 5481$\pm$70~K, $\log g$ of 3.83$\pm$0.10, and a metallicity of --0.11$\pm$0.07, the latter was revised to --0.09$\pm$0.07 from a single BESO spectrum (Fuhrmann et al. \cite{beso}). Its $v\sin i$ of 1.0 was just mentioned in a figure caption and it is not clear whether this was actually derived or assumed. Jofr\'e et al. (\cite{jof1}) lists results from an ELODIE spectrum with 5559\,K and a $\log g$ of 4.05. Our values are 5475\,K, $\log g$ of 3.86, and with a metallicity of --0.13. A rotational broadening was actually not detected.

$\alpha$~Tau's stellar parameters were determined by many sources summarized in Heiter et al. (\cite{heiter}). Its $T_{\rm eff}$ ranges between 3987 and 3887\,K, its $\log g$ between 1.20 and 1.42. Our values are 3900\,K, $\log g$ of 1.45, and a metallicity of --0.33. Continuum setting for $\alpha$~Tau was iteratively improved but many unknown lines make the fit vulnerable to line-list deficiencies. Also note that the $v\sin i$ of 3.5\,\kms\ from our $R$=220,000 spectra is smaller than any other value published so far. Otherwise, the benchmark values are matched properly.

\subsection{Rare-earth elements}

\begin{figure}
\includegraphics[angle=0,width=87mm]{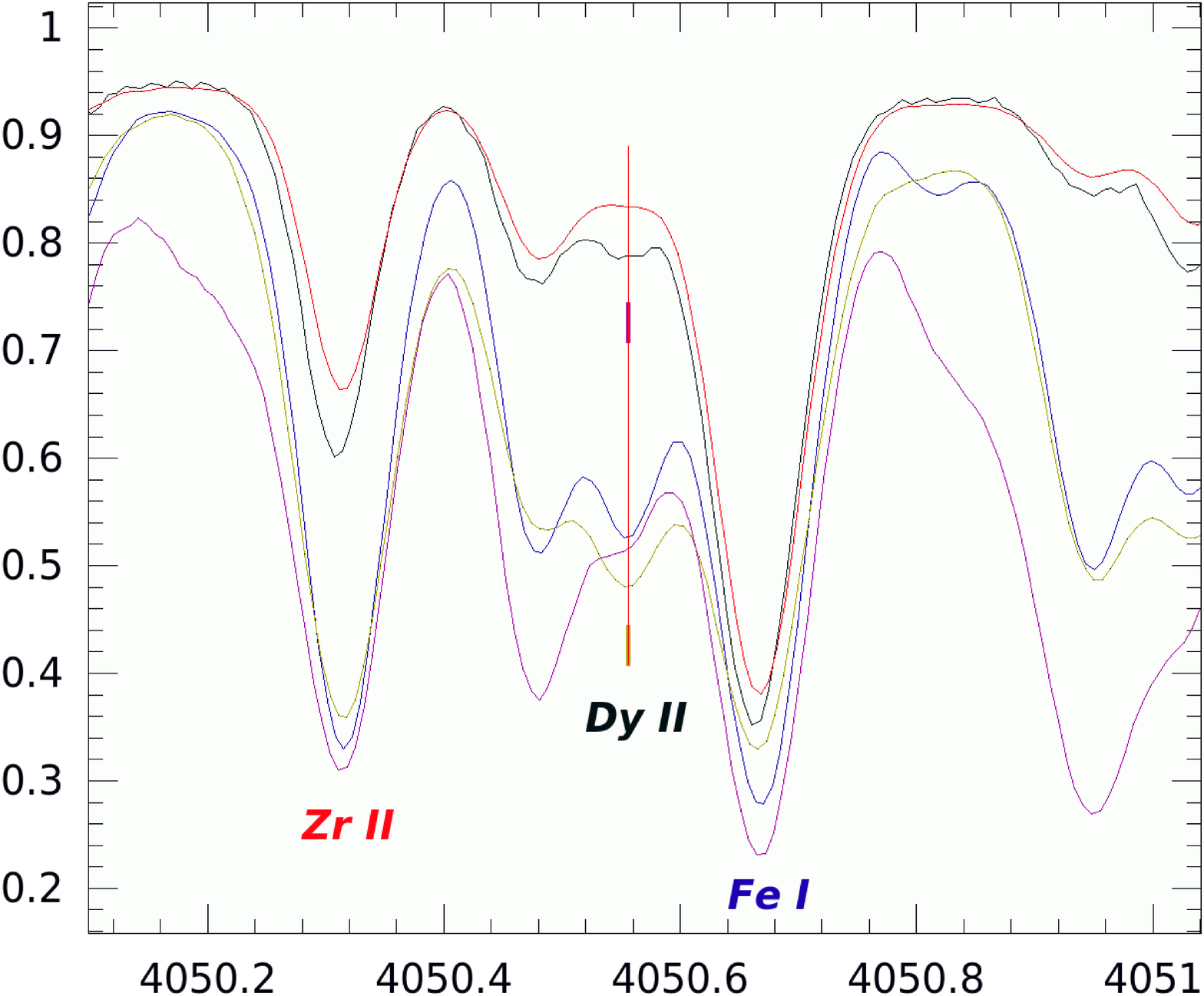}
\caption{Detection of the rare-earth element dysprosium (Dy) in the spectra of several library stars (from top to bottom for the Fe\,{\sc i} line core; Sun, 70\,Vir, $\alpha$\,Boo, $\alpha$\,Ari, $\mu$~Leo). Dysprosium's name comes from greek dusprositos and means ``hard to get at''. The singly-ionized Dy line at 4050.55\,\AA\ is indicated along with one Zr\,{\sc ii} and one Fe\,{\sc i} line. }
 \label{F-dys}
\end{figure}

Fig.~\ref{F-dys} shows the detection of the rare-earth element dysprosium ($Z$=66). The figure plots several stars ranging from main-sequence stars like the Sun and 70\,Vir to the RGB stars $\alpha$\,Boo, $\alpha$\,Ari, and $\mu$~Leo (all $\approx$K1-2\,III). The singly-ionized Dy line at 4050.55\,\AA\ is indicated with a vertical line. Dysprosium's name comes from greek dusprositos and means ``hard to get at'', which spurred us to use it as a case example. The Dy\,{\sc ii} line shown is weak and almost buried by a nearby Zr\,{\sc ii}, an Fe\,{\sc i}, and an unidentified line blend, but still among the more easily detectable lines. There are many more Dy lines in the spectral range of PEPSI that could be exploited for an abundance determination. The element has been detected in the solar spectrum as well as in some Ap stars (e.g., Ryabchikova et al. \cite{ryab}) and is possibly synthesized only in supernovae. Experimental wavelengths and oscillator strengths for Dy\,{\sc ii} lines are available from Wickliffe et al. (\cite{wick}).

\subsection{Carbon $^{12}$C to $^{13}$C ratio}

Mel\'endez et al. (\cite{mel09}) have shown that the Sun exhibits a depletion of refractory elements relative to volatile elements and related this to a cleansing effect during its rocky-planet formation. The engulfment of a close-orbiting planet during the RGB evolution may lead to the opposite effect, an overabundance of refractory elements, if fully transferred to the envelope of the host star. Various authors searched for signatures of such a replenishment in the Li abundances and its isotope ratio (e.g., Kumar et al.~\cite{kumar}) but also in $^{12}$C\,/\,$^{13}$C (e.g., Carlberg et al. \cite{carl}). Accurate C-isotope ratios require high-resolution and high S/N spectra while applicable only to stars cool enough to enable CN formation and with low $v\sin i$ (e.g., Berdyugina \& Savanov \cite{ber:sav}). No conclusive statements on engulfing could be made so far from carbon abundances but it seems clear that initial protostellar Li abundances and $^{12}$C\,/\,$^{13}$C may be more diverse than originally thought (Carlberg et al. \cite{carl}). It is thus advisable also analyzing the carbon isotope ratio in benchmark RGB stars.

\begin{figure}
\includegraphics[angle=0,width=87mm,clip]{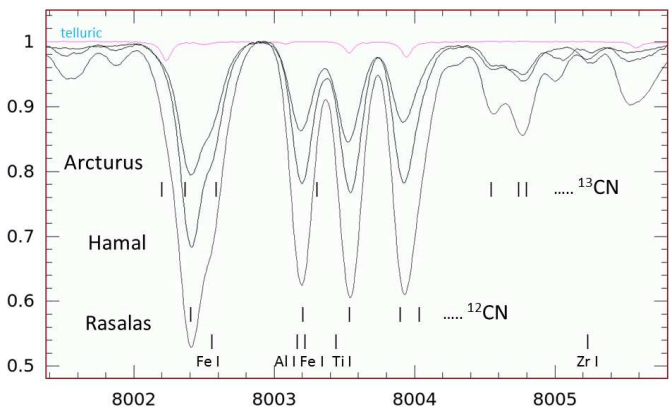}
\caption{Comparison of the 800-nm region of three RGB stars in this library; from top to bottom, Arcturus ($\alpha$\,Boo, K1.5III), Hamal ($\alpha$\,Ari, K1IIIb), Rasalas ($\mu$\,Leo, K2III). The region contains many $^{12}$CN and $^{13}$CN lines from which the $^{12}$C to $^{13}$C ratio is derived. A telluric spectrum scaled to the Rasalas observation is shown on the top. }
 \label{F-cratio}
\end{figure}

Fig.~\ref{F-cratio} shows a comparison of the 800-nm region of three of the RGB stars in this paper (Arcturus, K1.5III; Hamal, K1IIIb; and Rasalas, K2III). The measurement of $^{12}$C\,/\,$^{13}$C usually comes from fitting a small group of CN lines in the spectral region between 8001 and 8005\,\AA . There, one can see the isotope ratio even by eye, that is, by comparing the 8004\,\AA\ line depth, which is solely due to $^{12}$CN, to the depth of the feature at 8004.7\,\AA , which is solely due to $^{13}$CN. CN identifications in the plot were taken from Carlberg et al. (\cite{carl}). Spectra of this region are contaminated by telluric lines. We have demonstrated this in paper~I for the solar spectra and for the solar twin 18~Sco. Fig.~\ref{F-cratio} shows for comparison the Kurucz et al. (\cite{kur:fur}) telluric spectrum. For our plot the telluric spectrum was scaled and shifted to match the observed telluric spectrum of $\mu$\,Leo (and can thus only be compared to the $\mu$\,Leo spectrum in the plot in Fig.~\ref{F-cratio}). It is self explaining that one should take its contamination into account when determining the carbon isotope ratio.

\subsection{Heavy elements}

The formation of very heavy elements like uranium ($Z$=92) or thorium ($Z$=90) requires a process with rapid neutron capture, the so-called r-process, which is naturally happening during supernova explosions. The explosion ejects these elements into space where they are then available for the next generation of stars. Common uranium isotopes have half-life decay times of 10$^8$ to 10$^9$ years, and had been used to determine the age of the universe by observing ultra-metal-poor stars (Frebel et al. \cite{frebel}). Gopka et al. (\cite{gopka07}) identified Th\,{\sc ii} 5989.045\,\AA\ in the KPNO Arcturus spectrum from Hinkle et al. (\cite{hinkle}). They followed this up by the identification of more such lines from the KPNO atlas (Gopka et al. \cite{gopka13}). The most widely used line is Th\,{\sc ii} 4019.129\,\AA . Together with the nearby neodymium line Nd\,{\sc ii} 4018.8\,\AA\ this line is used to form the chronometric ratio Th/Nd, first suggested by Butcher (\cite{but}) as an age indicator. However, there is no systematic identification of heavy-element lines in benchmark stars.

\begin{figure}
\includegraphics[angle=0,width=87mm,clip]{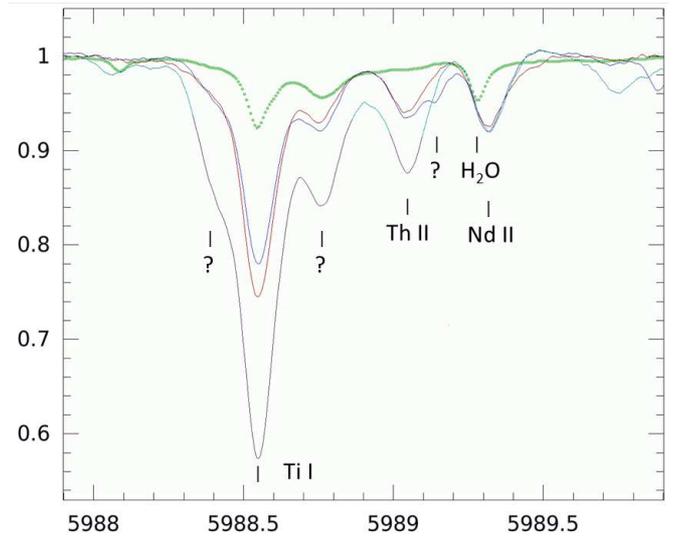}
\caption{Th\,{\sc ii} 5989.045-\AA\ and Nd\,{\sc ii} 5989.378-\AA\ lines in three RGB stars (full lines, from top to bottom in the Ti\,{\sc i} line core; Arcturus, Hamal, Rasalas). The top spectrum shown as dots is a spectrum of the Sun. }
 \label{F7}
\end{figure}

We have visually searched our library sample for the strong lines of U\,{\sc ii} at, for example 3859.57\,\AA\ and 4241.67\,\AA\ as well as Th\,{\sc ii} at 4019.129\,\AA\ and 5989.045\,\AA , leaning on new oscillator strengths from Nilsson  et al. (\cite{nil-Th}, \cite{nil-U}). Some of these lines are indeed identified, for example the Th\,{\sc ii}-Nd\,{\sc ii} line pair at 5989\,\AA\ in giant stars like Arcturus, $\alpha$\,Ari, or $\mu$\,Leo or the U\,{\sc ii} transition at 4241-\AA\ in $\epsilon$\,Vir. Fig.~\ref{F7} is a comparison of above three targets with the Sun for the Th 5989-\AA\ region. For many others there is only vague evidence due to blending. Note again that the solar spectrum in this plot was also taken with PEPSI in the same manner as the other stars. It shows a dominant telluric blend on the blue side of the expected Nd\,{\sc ii} line. Th\,{\sc ii} 5989.045 may be present in the solar spectrum with an upper limit at the $<$1\,m\AA\ level.

In connection with chronometry, the s-process elements Ba and Sr (a.o.) are also key elements. The most prominent barium line is the one Ba\,{\sc ii} transition at 4554.02\,\AA . Other comparably weaker lines are 3891.78\,\AA\ and 4130.64\,\AA\ (see, e.g., Siqueira Mello et al. \cite{sig:hil}). Strontium lines are comparably more prominent than U and Th, for example Sr\,{\sc ii} at 4077.72\,\AA\ or 5215.52\,\AA . The 4077-\AA\ line is commonly used in the M-K system to classify cool, low-gravity, stellar atmospheres because of its density sensitivity (Gray \& Garrison \cite{gra:gar}). Our library spectra may be used to investigate spectral lines from heavy elements in greater detail than before.

\section{Summary}\label{S7}

We present a library of high-resolution (on average $\approx$220,000), high-S/N optical spectra of a sample of 48 bright reference stars. The sample includes the northern \emph{Gaia} benchmark stars and a few well-known M-K standards. Deep spectra are build by average combining individual exposures and reach S/N of many hundreds and, in some cases, even thousands. Continua are set by predetermining synthetic spectra that match the target classification. Continuum adjustments of several tens of per cent were necessary for wavelengths shorter than 400\,nm. For all other wavelengths the differences in the continuum were always less than 1\%\ and for wavelength regions with clear continuum visibility more like 0.2\%.

A comparison with the KPNO Arcturus atlas revealed nearly perfect agreement with our library spectrum. At some wavelength regions the pixel-to-pixel differences are equal to or comparable to the photon noise level after both spectra were shifted in wavelength to the barycentric rest frame. We also found very weak \ion{Ca}{ii} H\&K and \Halpha\ residual emission in Arcturus, thereby further strengthening earlier claims that it has a magnetic field and even an H\&K activity cycle. The spectrum is still to be examined in detail whether the match also holds for other spectral lines. After all, the two spectra were taken 16\,yrs apart.

We identified several archival science cases possible to be followed up with the present data. Among these are the determination of global stellar parameters like effective temperature, gravity, metallicity, and elemental abundances. For a demonstration, we applied our spectrum synthesis code ParSES to a number of selected wavelength regions of 70~Vir (G4V-IV) and $\alpha$~Tau (K5III). The resulting values are summarized and compared with the literature in Table~\ref{T-parses}. These numbers  are not intended to be the final verdict but shall just demonstrate the capabilities and the expected uncertainties.

Of particular interest are isotopic line ratios. The most asked for in the literature is the $^6$Li to $^7$Li ratio from the two Li doublets at 6708\,\AA . Its science cases range from rocky planet engulfment, internal stellar mixing and dredge-up mechanisms, to the primordial Li production rate. In our paper~I on the Sun-as-a-star, we had analyzed this wavelength region of the Sun in detail and refer the reader to this paper. Another isotope ratio of general interest is $^{12}$C\,/\,$^{13}$C. Its primary science case is the main chain of the CNO cycle in stellar evolution but also allows the quantification of dredge-up episodes on the RGB in more detail. Finally, elemental abundances of species ``that are hard to get at'' are made accessible, for example the rare-earth element dysprosium or the heavy elements uranium and thorium, just to name a few.

The reduced deep spectra can be downloaded in FITS format from our web page at
\begin{center}
https://pepsi.aip.de \ .
\end{center}
We also provide the deep 1D data prior to the final continuum normalization.

\acknowledgements{We thank all engineers and technicians involved in PEPSI, in particular our Forschungstechnik team and its late Emil Popow who passed away much too early, but also Mark Wagner and John Little and his LBTO mountain crew. Our thanks also go to Christian Veillet, LBT director, and to Paul Gabor, VATT director, for their clever telescope scheduling. Walter Seifert, LSW Heidelberg, is warmly thanked for making some hours of his LUCI commissioning time available to us. It is also our pleasure to thank the German Federal Ministry (BMBF) for the year-long support through their Verbundforschung for LBT/PEPSI through grants 05AL2BA1/3 and 05A08BAC. Finally, the many telescope operators are thanked for their patience and even --partial-- enthusiasm when we wanted to observe Sirius with a 12m telescope. This research has made use of the SIMBAD database, operated at CDS, Strasbourg, France.}

\appendix

\section{Detailed observing log}

\begin{table*}
\caption{Log of individual spectra. } \label{AT1}
\begin{tabular}{lllllllllll}
\hline \noalign{\smallskip}
Star & MK &  \multicolumn{6}{c}{Individual-spectrum S/N} & $N$ & Date & Telescope\\
     &    &  I/404 & II/450 & III/508 & IV/584 & V/685 & VI/825 & & & \\
\noalign{\smallskip}\hline \noalign{\smallskip}
\emph{Giants} & & & & & & & & & & \\
\noalign{\smallskip}
32 Gem          & A9 III&100 &115 &205 &270 &265 &240 & 222 222 & 15Apr 10 & LBT \\
HD 140283       & F3 IV &140 &210 &290 &450 &480 &490 & 222 224 & 15May 25 & LBT \\
HD 122563       & F8 IV &100 &160 &245 &310 &405 &400 & 222 224 & 15Apr 10 & LBT \\
$\eta$ Boo      & G0 IV &400 &440 &600 &800 &750 &450 & 233 36(10) & 15Apr 9 & LBT \\
$\zeta$ Her     & G0 IV &\dots &\dots &260 &\dots &640 &\dots & 003 030  & 16Apr 2 & VATT \\
$\zeta$ Her     & G0 IV &\dots &120 &260 &395 &600 &530 & 032 222  & 16Apr 3 & VATT \\
$\zeta$ Her     & G0 IV &\dots &140 &295 &\dots &650 &540 & 011 011  & 16Apr 5 & VATT \\
$\zeta$ Her     & G0 IV &\dots &35 &\dots &170 &\dots &110 & 020 101  & 16Apr 7 & VATT \\
$\zeta$ Her     & G0 IV &\dots &70 &\dots &250 &\dots &\dots & 020 200  & 16Apr 10 & VATT \\
$\zeta$ Her     & G0 IV &690 & \dots & \dots & \dots & \dots &700 & 200 00(10)  & 16Jun 3& LBT \\
$\delta$ CrB    & G3.5III& 320 & 320 & 420 & 500 & 500 & 630 & 122 223 & 17Mar 3 & LBT \\
$\mu$ Her       & G5 IV &\dots &\dots &110 &\dots &200 &\dots & 002 020 & 15May 29 & VATT \\
$\mu$ Her       & G5 IV &\dots &50 &120 &190 &310 &300 & 012 111 & 15May 30 & VATT \\
$\mu$ Her       & G5 IV &\dots &40 &95 &145 &245 &250 & 012 111 & 15Jun 1 & VATT \\
$\mu$ Her       & G5 IV &\dots &50 &100 &160 &265 &255 & 012 111 & 15Jun 2 & VATT \\
$\mu$ Her       & G5 IV &570 &440 &480 &740 &710 &610 & 255 55(10) & 16Jun 3& LBT \\
$\beta$ Boo     & G8 III &\dots &300 &395 &500 &600 & \dots & 0(10)(10) (10)(10)0 & 16Jun 4& LBT \\
$\beta$ Boo     & G8 III &400 & \dots&\dots &\dots &\dots &550 & 300 00(14) & 16Jun 5& LBT \\
$\epsilon$ Vir  & G8 III &230 & \dots&\dots &\dots &\dots &500 & 300 008 & 15Apr 8 & LBT \\
$\epsilon$ Vir  & G8 III &\dots &70 &125 &185 &280 &330 & 063 333 & 15Apr 1 & VATT \\
$\beta$ Gem     & K0 IIIb &150 &\dots &\dots &\dots &\dots &330 & 600 008 & 15Apr 8 & LBT \\
$\beta$ Gem     & K0 IIIb &\dots &50 &185 &125 &425 &\dots & 033 330 & 15Apr 2 & VATT \\
$\beta$ Gem     & K0 IIIb &\dots &\dots &\dots &300  &\dots &\dots & 000 200 & 15Apr 3 & VATT \\
$\beta$ Gem     & K0 IIIb &\dots &70 &\dots &230 &\dots &500 & 060 303 & 15Apr 4 & VATT \\
HD 107328       & K0 IIIb &190 &250 &500 &650 &900 &990 & 343 436 & 15May 23 & LBT \\
$\alpha$ UMa    & K0 III &230 &\dots &\dots &\dots &\dots &370 & 300 006 & 15Apr 9 & LBT \\
$\alpha$ UMa    & K0 III &\dots &75 &170 &300 &440 &400 & 063 333 & 15Apr 6 & VATT \\
$\alpha$ Ari    & K1 IIIb &700 &\dots &\dots &900 &\dots &\dots & 300 800 & 15Sep 27 & LBT \\
$\alpha$ Ari    & K1 IIIb &\dots &260 &300 &\dots &570 &650 & 066 066 & 15Nov. 20 & LBT \\
$\alpha$ Boo    & K1.5 III &250 &500 &\dots &350 &\dots &500 & 630 509 & 15Apr 8 & LBT \\
$\alpha$ Boo    & K1.5 III &\dots &105 &245 &420 &620 &\dots & 034 340 & 15Apr 2 & VATT \\
$\alpha$ Boo    & K1.5 III &\dots &110 &\dots &\dots &\dots &650 & 080 008 & 15Apr 3 & VATT \\
$\alpha$ Boo    & K1.5 III &60 &\dots &\dots &\dots &350 &\dots & 300 0(21)0 & 15Apr 5 & VATT \\
$\alpha$ Boo    & K1.5 III &75 &\dots &\dots &\dots &450 &\dots & 300 0(21)0 & 15Apr 6 & VATT \\
7 Psc           & K2 III &170 &295 &550 &660 &1000 &900 & 233 334 & 15Nov 20 & LBT\\
$\mu$ Leo       & K2 III &80 &\dots &\dots &\dots &\dots &300 & 200 00(10) & 15Apr 8 & LBT \\
$\mu$ Leo       & K2 III &200 &210 &300 &430 &530 &550 & 234 346 & 15Apr 9 & LBT \\
$\gamma$ Aql    & K3 II & \dots &\dots &85 &\dots &210 &\dots & 001 010 & 15May 7 & VATT \\
$\gamma$ Aql    & K3 II & \dots &150 &360 &570 &900 &845 & 021 111 & 16May 24 & VATT \\
$\gamma$ Aql    & K3 II & \dots &150 &330 &320 &755 &890 & 021 111 & 16May 26 & VATT \\
$\gamma$ Aql    & K3 II & \dots &150 &350 &440 &800 &935 & 021 111 & 16May 28 & VATT \\
$\gamma$ Aql    & K3 II & \dots &160 &360 &500 &870 &960 & 021 111 & 16May 29 & VATT \\
$\gamma$ Aql    & K3 II & \dots &100 &250 &315 &590 &340 & 021 111 & 16May 30 & VATT \\
$\gamma$ Aql    & K3 II  &410 & \dots & \dots & \dots & \dots &950 & 200 006 & 16Jun 5 & LBT \\
$\beta$ UMi     & K4 III &160 &200 &400 &560 &900 &750 & 433 33(12) & 15May 23 & LBT \\
$\beta$ UMi     & K4 III &\dots &\dots &50 &\dots &150 &\dots & 003 0(21)0 & 15Apr 7 & VATT \\
$\alpha$ Tau    & K5 III &\dots &180 &210 &695 &650 &1200 & 063 335 & 15Sep 19 & VATT \\
$\alpha$ Tau    & K5 III &480 &500 &920 &600 &1300 &1400 & 355 955 & 16Oct 1 & LBT \\
$\mu$ UMa       & M0 III &\dots &50 &80 &190 &300 &300 & 064 343 & 15Apr 4 & VATT \\
$\mu$ UMa       & M0 III &160 &\dots &\dots &\dots &\dots &400 & 300 00(12) & 15Apr 9 & LBT \\
$\mu$ UMa       & M0 III &\dots &140 &340 &420 &730 &\dots & 032 420 & 17Mar 3 & LBT \\
$\gamma$ Sge    & M0 III &\dots &50 &100 &210 &355 &\dots & 011 110 & 15May 27 & VATT \\
$\gamma$ Sge    & M0 III &\dots &\dots &145 &\dots &450 &\dots & 001 010 & 15May 28 & VATT \\
$\gamma$ Sge    & M0 III &\dots &40 &50 &205 &245 &\dots & 011 110 & 15May 29 & VATT \\
$\gamma$ Sge    & M0 III &\dots &50 &95 &200 &340 &\dots & 011 110 & 15May 30 & VATT \\
$\gamma$ Sge    & M0 III &\dots &40 &95 &200 &350 &\dots & 011 110 & 15Jun 1 & VATT \\
$\gamma$ Sge    & M0 III &\dots &40 &95 &200 &340 &\dots & 011 110 & 15Jun 2 & VATT \\
$\gamma$ Sge    & M0 III &\dots &60 &110 &290 &335 &525 & 041 222 & 15Sep 19 & VATT \\
$\gamma$ Sge    & M0 III &\dots &63 &120 &220 &310 &500 & 042 222 & 15Sep 21 & VATT \\
$\gamma$ Sge    & M0 III &\dots &\dots &260 & \dots &500 &\dots & 002 020 & 15Sep 25& LBT \\
$\gamma$ Sge    & M0 III &340 &\dots & \dots & \dots & \dots &850 & 200 00(13) & 16Jun 3& LBT \\
$\alpha$ Cet    & M1.5 IIIa &480 &\dots &\dots &900 &\dots &\dots & 300 900 & 15Sep 27 & LBT \\
$\alpha$ Cet    & M1.5 IIIa &\dots &190 &400 &\dots &750 &980 & 026 062 & 15Nov 20 & LBT \\
\noalign{\smallskip}\hline
\end{tabular}
\tablefoot{As for Table~\ref{T1}. $Date$ is the night of observation in UT (e.g., 16Apr means April 2016.}
\end{table*}

\setcounter{table}{1}
\begin{table*}
\caption{(continued) }
\begin{tabular}{lllllllllll}
\hline \noalign{\smallskip}
Star & MK &  \multicolumn{6}{c}{Individual-spectrum S/N} & $N$ & Date & Telescope\\
     &    &  I/404 & II/450 & III/508 & IV/584 & V/685 & VI/825 & & & \\
\noalign{\smallskip}\hline \noalign{\smallskip}
\noalign{\smallskip}
\emph{Dwarfs} & & & & & & & & & & \\
\noalign{\smallskip}
$\alpha$\,CMa   & A1 V &800 &590 &720 &740 &620 &420 & 111 111 & 16 Oct 1 & LBT \\
HD 84937        & F2 V &35 &75 &120   &110 &150 &75 & 222 222 & 15Apr 2 & LBT \\
HD 84937        & F2 V &66 &80 &\dots &100 &\dots &180 & 130 301 & 17Mar 3 & LBT \\
$\sigma$ Boo    & F4 V & \dots & \dots&165 &\dots &390 & \dots& 003 030 & 16Apr 2 & VATT \\
$\sigma$ Boo    & F4 V & \dots &85 &170 &235 &360 &295 & 021 111 & 16Apr 3 & VATT \\
$\sigma$ Boo    & F4 V & \dots &85 &180 &285 &380 &305 & 021 111 & 16Apr 4 & VATT \\
$\sigma$ Boo    & F4 V & \dots &60 &\dots &195 &\dots &\dots & 020 200 & 16Apr 10 & VATT \\
$\sigma$ Boo    & F4 V & \dots &115 &\dots &\dots &\dots &390 & 010 001 & 16May 25 & VATT \\
$\sigma$ Boo    & F4 V & \dots &100 &\dots &\dots &\dots &380 & 010 001 & 16May 27 & VATT \\
$\sigma$ Boo    & F4 V & \dots &\dots &220 &\dots &390 &\dots & 001 010 & 16May 29 & VATT \\
$\sigma$ Boo    & F4 V & \dots &115 &\dots &\dots &\dots &380 & 010 001 & 16May 31 & VATT \\
$\sigma$ Boo    & F4 V & \dots &102 &\dots &\dots &\dots &\dots & 010 000 & 16Jun 2 & VATT \\
$\sigma$ Boo    & F4 V &460 &\dots & \dots & \dots & \dots &650 & 200 006 & 16Jun 3 & LBT \\
HD 49933        & F5 V-IV &\dots &\dots &240 &\dots &400 &\dots & 004 040 & 16Apr 11 & LBT \\
HD 49933        & F5 V-IV &280 &320 &500 &600 &670 &510 & 122 222 & 16Oct 1 & LBT \\
$\alpha$ CMi    & F5 V-IV &\dots &120 &225 &350 &500 &430 & 0(12)6 666 & 15Apr 5 & VATT \\
$\alpha$ CMi    & F5 V-IV &280 &\dots &\dots &\dots &\dots &310 & 500 008 & 15Apr 8 & LBT \\
$\theta$ UMa    & F7 V & \dots &110 &215 &330 &485 &445 & 042 222  & 16Apr 4 & VATT \\
$\theta$ UMa    & F7 V & \dots &80 &160 &230 &300 &300 & 042 222  & 16Apr 5 & VATT \\
$\theta$ UMa    & F7 V & \dots &80 &200 &220 &430 &360 & 042 222  & 16Apr 6 & VATT \\
$\theta$ UMa    & F7 V & \dots &96 &170 &265 &390 &\dots & 022 220  & 16Apr 10 & VATT \\
$\theta$ UMa    & F7 V &120  & \dots & \dots& \dots& \dots& 260 & 100 001  & 16Apr 11& LBT \\
$\theta$ UMa    & F7 V &450  & \dots & \dots& \dots& \dots& 500 & 200 006  & 16Jun 3& LBT \\
$\beta$ Vir     & F9 V &225 &\dots &\dots &\dots &\dots &350 & 200 007 & 15Apr 8 & LBT \\
$\beta$ Vir     & F9 V & \dots &55 &95 &120 &175 &\dots & 043 430  & 15Apr 3 & VATT \\
$\beta$ Vir     & F9 V & \dots &50 &\dots &\dots &\dots &160 & 030 003 & 15Apr 5 & VATT \\
HD 22879        & F9 V &\dots &330 &410 &590 &570 &\dots & 021 210 & 15Sep 27 & LBT \\
HD 22879        & F9 V &205 &\dots &\dots &\dots &\dots &420 & 100 001 & 16Oct 1 & LBT \\
HD 189333       & F9 V &88 &170 &245 &230 &290 &340 & 111 111 & 16Oct 1 & LBT \\
HD 52265        & G0 V & \dots &\dots &97 &\dots &168 &80 & 001 011 & 16Apr 5 & VATT \\
HD 52265        & G0 V & \dots & &30 &\dots &76 &50 & 011 011 & 16Apr 6 & VATT \\
HD 159222       & G1V  &210 &260 &375 &500 &560 &530 & 232 323 & 16Jun 5 & LBT\\
16 Cyg A        & G1.5 V &180 &215 &410 &540 &630 &500 & 122 223 & 15May 25 & LBT \\
Sun             & G2 V & & & & & & & 444 446 & 15Sep 24 & SDI \\
HD 101364       & G2 V &\dots &\dots &80 &\dots &110 & \dots & 003 030 & 15Apr 8 & LBT \\
HD 101364       & G2 V &55 &110 &145 &185 &200 &185 & 333 333 & 15Apr 10 & LBT \\
HD 82943        & G2 V &50 &120 &160 &200 &220 &160 & 322 223 & 15Apr 2 & LBT \\
18 Sco          & G2 V &180 &250 &450 &550 &700 &690 & 233 333 & 15May 23 & LBT \\
18 Sco          & G2 V &270 &\dots &\dots &450 &590 &\dots & 200 440& 15May 24 & LBT \\
51 Peg          & G2.5 V &240 &380 &550 &640 &810 &700 & 344 445 & 15Nov 20& LBT \\
16 Cyg B        & G3 V &\dots &390 &420 &650 &650 &650 & 022 223 & 15May 25 & LBT \\
16 Cyg B        & G3 V &220 &\dots &\dots &\dots &670 &\dots & 100 040 & 15Sep 27 & LBT \\
70 Vir          & G4 V &250 &380 &680 &810 &1000 &900 & 222 224 & 15May 25 & LBT \\
$\mu$ Cas       & G5 V &480 &420 &700 &810 &1080 &1100 & 122 222 & 15Sep 27 & LBT \\
HD 103095       & G8 V &105 &145 &220 &275 &325 &325 & 233 433 & 15Apr 10 & LBT \\
$\tau$ Cet      & G8.5 V &\dots &66 &165 &\dots &385 &380 & 011 011 & 15Sep 20 & VATT \\
$\tau$ Cet      & G8.5 V &450 &\dots &\dots &540 &700 &\dots & 400 380 & 15Sep 27 & LBT \\
$\tau$ Cet      & G8.5 V &\dots &\dots &184 &\dots &390 &\dots & 001 010 & 15Nov 28 & VATT \\
$\tau$ Cet      & G8.5 V &\dots &\dots &150 &\dots &400 &\dots & 001 010 & 15Nov 29 & VATT \\
$\tau$ Cet      & G8.5 V &\dots &\dots &138 &\dots &398 &\dots & 001 010 & 15Nov 30 & VATT \\
$\tau$ Cet      & G8.5 V &\dots &\dots &191 &\dots &537 &\dots & 001 010 & 15Dec 1 & VATT \\
$\tau$ Cet      & G8.5 V &\dots &\dots &153 &\dots &435 &\dots & 001 010 & 15Dec 2 & VATT \\
$\tau$ Cet      & G8.5 V &\dots &\dots &230 &\dots &580 &\dots & 001 010 & 15Dec 3 & VATT \\
$\tau$ Cet      & G8.5 V &\dots &\dots &225 &\dots &574 &\dots & 001 010 & 15Dec 4 & VATT \\
$\tau$ Cet      & G8.5 V &\dots &\dots &101 &\dots &237 &\dots & 001 010 & 15Dec 5 & VATT \\
$\tau$ Cet      & G8.5 V &\dots &\dots &210 &\dots &520 &\dots & 001 010 & 15Dec 6 & VATT \\
$\tau$ Cet      & G8.5 V &\dots &370 &530 &750 &\dots &710 & 055 505 & 16Oct 1 & LBT \\
$\epsilon$ Eri  & K2 V & \dots &\dots &185 &\dots &475 &\dots & 001 010 & 15Sep 24 & VATT \\
$\epsilon$ Eri  & K2 V &540 &450 &630 &800 &1100 &1000 & 155 555 & 16Oct 1 & LBT \\
HD 192263       & K2 V &172 &209 &310 &330 &450 &480 & 122 222 & 16Jun 6 & LBT \\
HD 128311       & K3 V &45 &100 &130 &180 &210 &200 & 322 223 & 15Apr 8 & LBT \\
HD 82106        & K3 V &50 &80 &140 &195 &220 &210 & 213 132 & 15Apr 1 & LBT \\
61 Cyg A        & K5 V &\dots &\dots & &\dots & &\dots & 002 020 & 15May 25 & LBT \\
61 Cyg A        & K5 V &240 &520 &700 &900 &1000 &850 & 122 334 & 15Sep 27 & LBT \\
61 Cyg B        & K7 V &176 &230 &440 &550 &910 &720 & 122 224 & 15Sep 27 & LBT \\
\noalign{\smallskip}\hline
\end{tabular}
\end{table*}

\end{document}